\documentclass[twocolumn,doi]{aastex63}

\newcommand{\MgIIdblt}{{\rm Mg}\kern 0.1em{\sc ii}~$\lambda\lambda 2796, 2803$}
\newcommand{\MgI}{\hbox{{\rm Mg}\kern 0.1em{\sc i}}}
\newcommand{\MgII}{\hbox{{\rm Mg}\kern 0.1em{\sc ii}}}
\newcommand{\MnII}{\hbox{{\rm Mn}\kern 0.1em{\sc ii}}}
\newcommand{\FeII}{\hbox{{\rm Fe}\kern 0.1em{\sc ii}}}
\newcommand{\OVIdblt}{{\rm O}\kern 0.1em{\sc vi}~$\lambda\lambda 1031, 1037$} 
\newcommand{\OVI}{\hbox{{\rm O}\kern 0.1em{\sc vi}}}
\newcommand{\CII}{\hbox{{\rm C}\kern 0.1em{\sc ii}}}
\newcommand{\CIV}{\hbox{{\rm C}\kern 0.1em{\sc iv}}}
\newcommand{\HI}{\hbox{{\rm H}\kern 0.1em{\sc i}}}
\newcommand{\Lya}{\hbox{{\rm Ly}\kern 0.1em$\alpha$}}
\newcommand{\Ha}{\hbox{{\rm H}\kern 0.1em$\alpha$}}
\newcommand{\Lyb}{\hbox{{\rm Ly}\kern 0.1em$\beta$}}
\newcommand{\OI}{\hbox{{\rm O}\kern 0.1em{\sc i}}}
\newcommand{\SiII}{\hbox{{\rm Si}\kern 0.1em{\sc ii}}}
\newcommand{\SiIII}{\hbox{{\rm Si}\kern 0.1em{\sc iii}}}
\newcommand{\SiIV}{\hbox{{\rm Si}\kern 0.1em{\sc iv}}}
\newcommand{\NII}{\hbox{{\rm N}\kern 0.1em{\sc ii}}}
\newcommand{\NV}{\hbox{{\rm N}\kern 0.1em{\sc v}}}
\newcommand{\kms}{\hbox{km~s$^{-1}$}}
\newcommand{\cmsq}{\hbox{cm$^{-2}$}}
\newcommand{\etal}{et~al.}

\accepted{October 26, 2020}
\submitjournal{ApJ}

\shorttitle{Galaxy Outflows at $z=2.071$ with KCWI}
\shortauthors{Nielsen {\etal}}

\begin{document}

\title{The CGM at Cosmic Noon with KCWI: Outflows from a Star-forming
  Galaxy at $z=2.071$}

\correspondingauthor{Nikole M. Nielsen}
\email{nikolenielsen@swin.edu.au}

\author[0000-0003-2377-8352]{Nikole M. Nielsen}
\affil{Centre for Astrophysics and Supercomputing, Swinburne
  University of Technology, Hawthorn, Victoria 3122, Australia}
\affil{ARC Centre of Excellence for All Sky Astrophysics in 3
  Dimensions (ASTRO 3D)}

\author[0000-0003-1362-9302]{Glenn G. Kacprzak}
\affil{Centre for Astrophysics and Supercomputing, Swinburne
  University of Technology, Hawthorn, Victoria 3122, Australia}
\affil{ARC Centre of Excellence for All Sky Astrophysics in 3
  Dimensions (ASTRO 3D)}

\author[0000-0002-3846-0980]{Stephanie K. Pointon}
\affil{Centre for Astrophysics and Supercomputing, Swinburne
  University of Technology, Hawthorn, Victoria 3122, Australia}
\affil{ARC Centre of Excellence for All Sky Astrophysics in 3
  Dimensions (ASTRO 3D)}

\author[0000-0002-7040-5498]{Michael T. Murphy}
\affil{Centre for Astrophysics and Supercomputing, Swinburne
  University of Technology, Hawthorn, Victoria 3122, Australia}

\author[0000-0002-9125-8159]{Christopher W. Churchill}
\affil{Department of Astronomy, New Mexico State University, Las
  Cruces, NM, 88003, USA}

\author[0000-0003-2842-9434]{Romeel Dav\'{e}}
\affil{Institute for Astronomy, Royal Observatory, University of
  Edinburgh, Edinburgh EH9 3HJ, UK}

\begin{abstract}

  We present the first results from our CGM at Cosmic Noon with KCWI
  program to study gas flows in the circumgalactic medium (CGM) at
  $z=2-3$. Combining the power of a high-resolution VLT/UVES quasar
  spectrum, an {\it HST}/ACS image, and integral field spectroscopy
  with Keck/KCWI, we detected {\Lya} emission from a $1.7L_{\ast}$
  galaxy at $z_{\rm gal}=2.0711$ associated with a Lyman limit system
  with weak {\MgII} ($W_r(2796)=0.24$~{\AA}) in quasar field
  J143040$+$014939. The galaxy is star-forming (${\rm SFR}_{\rm
    FUV}=37.8$~M$_{\odot}$~yr$^{-1}$) and clumpy: either an edge-on
  disk ($i=85^{\circ}$) or, less likely, a major merger. The
  background quasar probes the galaxy at an impact parameter of
  $D=66$~kpc along the projected galaxy minor axis
  ($\Phi=89^{\circ}$). From photoionization modeling of the absorption
  system, we infer a total line-of-sight CGM metallicity of ${\rm
    [Si/H]}=-1.5^{+0.4}_{-0.3}$. The absorption system is roughly
  kinematically symmetric about $z_{\rm gal}$, with a full {\MgII}
  velocity spread of $\sim210$~{\kms}. Given the galaxy--quasar
  orientation, CGM metallicity, and gas kinematics, we interpret this
  gas as an outflow that has likely swept-up additional material. By
  modeling the absorption as a polar outflow cone, we find the gas is
  decelerating with average radial velocity $V_{\rm
    out}=109-588$~{\kms} for half opening angles of
  $\theta_0=14^{\circ}-75^{\circ}$. Assuming a constant $V_{\rm out}$,
  it would take on average $t_{\rm out}\sim111-597$~Myr for the gas to
  reach 66~kpc. The outflow is energetic, with a mass outflow rate of
  $\dot{M}_{\rm out}<52{\pm37}$~M$_{\odot}$~yr$^{-1}$ and mass loading
  factor of $\eta<1.4{\pm1.0}$. We aim to build a sample of $\sim50$
  {\MgII} absorber--galaxy pairs at this epoch to better understand
  gas flows when they are most actively building galaxies.

\end{abstract}

\keywords{Galaxy evolution (594), High-redshift galaxies (734),
  Lyman-alpha galaxies (978), Quasar absorption line spectroscopy
  (1317), Circumgalactic medium (1879)}

\section{Introduction}
\label{sec:intro}

One of the most challenging problems for deciphering how galaxies
evolve is understanding how they obtain their gas and process it into
stars. Simulations and simple models suggest that the baryon cycle, or
the flow of gas onto, out of, and back onto galaxies, is the main
regulator of star formation \citep[e.g.,][]{oppenheimer08,
  lilly-bathtub}. We further know that the cosmic star formation rate
peaks at $z=2-3$, also known as ``Cosmic Noon,'' when galaxies
assemble roughly half of their stellar mass \citep{madau14}. At this
epoch, strong galactic outflows are regularly observed
\citep[e.g.,][]{steidel10, rupke-review} and simulations predict the
rate of accretion is greatest \citep[e.g.,][]{vandevoort11b}. As a
result, Cosmic Noon is the ideal epoch for studying the baryon
cycle. The ideal location for this study is the circumgalactic medium
\citep[CGM;][]{tumlinson17}, which is generally defined as the bound
gaseous halo around galaxies. Accreting gas from the intergalactic
medium (IGM) to the interstellar medium (ISM) must first pass through
the CGM, which is also where most outflowing gas is deposited. The CGM
is therefore the record-keeper of past gas flows and a reservoir for
future star formation.

At low redshift, $z\lesssim1$, we now understand that the CGM is
massive, containing a gas mass at least comparable to the gas mass
within galaxies themselves \citep{thom11, tumlinson11, werk13}, and
holds a substantial fraction of baryons \citep[e.g.,][]{peeples14,
  werk14}. Quasar absorption line spectroscopy has shown that this
multiphase gas is accreting/rotating \citep[e.g.,][]{ggk-sims,
  kacprzak19a, magiicat5, ho17, ho20, zabl19} and outflowing
\citep[e.g.,][]{kacprzak14, kacprzak19a, muzahid15, magiicat5,
  schroetter16, schroetter19, martin19}. These gas flows are likely
confined to the projected major and minor axes of their host galaxies,
respectively \citep{bordoloi11, bouche12, kcn12, kacprzak15, lan14,
  lan18, schroetter19}, and there is evidence that their metallicities
are bimodal \citep[at least for partial Lyman limit systems and Lyman
  limit systems, pLLSs/LLSs, at $0.45<z<1.0$;][]{lehner13, lehner19,
  wotta16, wotta19}. One might then assume that there should be two
clear populations of gas on the azimuthal angle--CGM metallicity
plane. Recent work by \citet{pointon19iso} suggests that this simple
model of metal-poor accretion along the projected major axis and
metal-rich outflows along the projected minor axis is not clearly
observed. Furthermore, comparing the CGM metallicity to the galaxy ISM
metallicity does not seem to improve the situation, where the gas flow
metallicities still have a wide range of values compared to the host
galaxies regardless of the azimuthal angle \citep{peroux16,
  kacprzak19b}. It is possible that this experiment simply needs a
larger sample to detect this bimodality \citep[e.g.,][]{peroux20} or
the metallicity analysis needs to account for dust
\citep[e.g.,][]{wendt20}. Alternatively, since gas flows are likely
diminishing in strength from Cosmic Noon towards present day, we might
expect their signatures to be weaker and more well-mixed at low $z$,
resulting in a lack of a metallicity difference for the observed gas
flows (\citealp[e.g.,][]{hafen17, hafen19}; although
\citealp{peroux20} suggest that there may not be a
metallicity--azimuthal angle bimodality at high redshift). The next
frontier for this sort of investigation is then high redshift.

Up until recently obtaining a large sample of absorber--galaxy pairs
was a time-intensive activity. Now the newest powerful method for
studying the CGM in detail is integral field spectroscopy. VLT/MUSE
campaigns are quickly building large samples of absorber--galaxy pairs
with a field-of-view of 1\arcsec,~covering a wide redshift range below
$z<1.5$ and above $z>3$. At low redshift, these campaigns find
galaxies hosting CGM absorption \citep[e.g.,][]{schroetter19,
  dutta20}, detect faint galaxies near CGM host galaxies that were
previously thought to be isolated \citep[e.g.,][]{peroux17, rahmani18,
  hamanowicz20}, easily obtain two dimensional information about all
galaxies in the fields for morphology and kinematic analyses
\citep[e.g.,][]{schroetter16, schroetter19, zabl19}, and allow for
multiple probes of the CGM hosted by a single galaxy using multiple
background galaxies \citep[e.g.,][]{peroux18} or even single
background lensed galaxies \citep[e.g., gravitational-arc
  tomography;][]{lopez18, lopez20}. At the moment, MUSE cannot reach
the epoch at which galaxies are most actively building up their mass,
$z=2-3$, because it does not yet have the spectral coverage required.

The Keck Baryonic Structure Survey \citep[KBSS;][]{rudie12} is
currently the largest survey of $z=2-3$ CGM absorber--galaxy pairs,
where they have obtained galaxy redshifts using multi-object
spectroscopy on Keck/MOSFIRE. By studying the redshift space
distortions around these galaxies, the KBSS have found enhancements of
metal lines out to 180~kpc and $\sim240-350$~{\kms}, which were
interpreted as inflows \citep{turner14, turner17} and hot,
metal-enriched outflows \citep{turner15}. Neutral hydrogen also has
significant enhancements in redshift space distortions, where
\citet{ychen20} stacked several thousand foreground--background galaxy
pairs and suggested that outflows dominate {\HI} kinematics up to
50~kpc from galaxies but inflows dominate beyond
100~kpc. \citet{rudie19} studied the multiphase CGM of eight galaxies,
finding substantial metal reservoirs ($>25\%$ of the ISM metal mass),
complex kinematics, unbounded gas, and high covering fractions
($>50\%$). These characteristics point to outflowing gas and a dynamic
CGM, where both heating and cooling processes occur often. To better
understand the source of the observed absorption, two key pieces of
information that are important for enhancing our insights are absent
from the KBSS: CGM metallicities and galaxy morphologies.

A few $z=2-3$ absorber--galaxy pairs have galaxy morphologies and CGM
metallicities. These include the damped {\Lya} (DLA) absorbers from
\citet{bouche13} and \citet{krogager13}. Both DLAs are metal-rich for
$z=2-3$ systems \citep[e.g.,][]{lehner16}, but are located along the
major and minor axes, respectively. \citet{bouche13} concluded that
the major axis DLA was consistent with a coplanar gaseous accretion
disk since the gas metallicity was significantly lower than the galaxy
ISM metallicity and the kinematics pointed to a combination of
accretion and rotating disk components. \citet{krogager13} concluded
that the minor axis DLA probed outflowing gas based on its location
about the galaxy and high metallicity, but did not model the
kinematics. An important caveat to these results is that both
absorber--galaxy pairs have very low impact parameters of 26 and
6~kpc, respectively, and are likely tracing some component of an
extended galaxy disk itself.

Even without galaxy morphologies, CGM metallicities have proven useful
for better constraining the gas flow origins at high
redshift. \citet{crighton15} studied a $z\sim2.5$ pLLS, finding
metal-enriched clouds with a large velocity width. The authors
concluded that the gas was consistent with an outflowing wind if they
assumed the host was an edge-on galaxy probed along the projected
minor axis. More confidently confirming this scenario requires
modeling the galaxy morphology and gas
kinematics. \citet{zahedy19accretion} found several {\Lya} emitters at
$z\sim2.8$ associated with strong {\Lya} absorption in a background
quasar spectrum with metallicities as low as $\sim 0.01-0.001$ solar,
which is consistent with the level of enrichment in the IGM at this
redshift. The authors suggested these low mass {\Lya} emitters were
therefore embedded in an IGM accretion stream.

Building up a large sample of absorber--galaxy pairs at Cosmic Noon
surrounding a variety of galaxies is important for studying these gas
flows during a highly influential epoch in the Universe's
history. Here we present the first results from our CGM at Cosmic Noon
with KCWI program. This paper is organized as
follows. Section~\ref{sec:methods} details the observations and
analysis methods for the quasar spectra, {\it HST} images, and IFU
spectroscopy. Section~\ref{sec:results} presents the CGM of a host
galaxy at $z\sim2$, characterizes the CGM metallicities and
kinematics, and explores an outflowing wind as the origin of the
observed gas. Section~\ref{sec:discussion} places the results in the
context of Cosmic Noon star formation-driven outflows and other
possible mechanisms that could give rise to the absorbing
gas. Finally, Section~\ref{sec:summary} summarizes and concludes our
first study with KCWI at Cosmic Noon. Throughout the paper we report
AB magnitudes, physical distances, and adopt a $\Lambda$CDM cosmology
($H_0=70$~{\kms}~Mpc$^{-1}$, $\Omega_M=0.3$, and
$\Omega_{\Lambda}=0.7$).

\section{Observations and Methods}
\label{sec:methods}

Our CGM at Cosmic Noon with KCWI program is designed to obtain
$\sim50$ {\MgII} absorber--galaxy pairs with the following
considerations. We compiled a sample of quasar fields with known
{\MgII} absorption at $1.9 < z_{\rm abs} < 2.6$, where each field has
been imaged with {\it HST} (ACS, WFPC2, and/or WFC3) or Keck (NIRC2
with laser guide star adaptive optics) and high-resolution quasar
spectra are available from VLT/UVES and/or HIRES/Keck. The quasar
spectra are sourced from KODIAQ DR1 \citep{kodiaqdr1}, SQUAD DR1
\citep{uvessquad}, or \citet[][also see
  \citealp{churchill20}]{evans-thesis}. The latter quasar sample
includes many quasars that are listed in KODIAQ DR2
\citep{kodiaqdr2}. The absorbers were drawn from the
\citet{mathes-thesis} and \citet{hasan20} sample, where {\MgII} and
      {\CIV} absorption were systematically and automatically searched
      using a matched-filter analysis similar to \citet{zhu13} and
      then later visually confirmed.

For each field in our program, multiple $z\sim2$ {\MgII} and {\CIV}
absorbers have been identified at $z_{\rm abs}<z_{\rm qso}-(1+z_{\rm
  qso})(3000$~{\kms}$/c)$ to avoid systems associated with the quasar.
We have placed no restrictions on the measured equivalent width of the
{\MgII} and {\CIV} lines other than the requirement that they be
detected at the $5\sigma$ level. Finally, we had no a priori knowledge
of the host galaxy properties (i.e., galaxies were not yet
identified). With this information in-hand, we have thus far observed
15 fields with the Keck Cosmic Web Imager \citep[KCWI;][]{kcwi} on
Keck II, corresponding to 30 {\MgII} and 80 {\CIV} absorbers,
searching for {\Lya} emission from galaxies at Cosmic Noon. While this
sample is initially {\MgII} absorption-selected, the IFU nature of
KCWI allows us to identify {\it all} $z\sim2$ galaxies within the
field of view of the instrument. We expect to identify at least as
many {\MgII} non-absorbing galaxies as {\MgII} absorbing galaxies. In
particular, a majority of the 80 {\CIV} absorbers we have covered in
the sample do not have measurable {\MgII} absorption.

Here we present the first result from our program: a {\MgII}
absorber--galaxy pair at $z=2.071$ in quasar field J143040$+$014939
(hereafter J1430$+$014).

\subsection{Quasar Spectroscopy and Photoionization Modeling}
\label{sec:qsospec}

Quasar J1430$+$014, $z_{\rm qso}=2.119$ was observed with VLT/UVES for
a total of 15077~s (PIDs 079.A-0656(A) and 081.A-0478(A)). The
spectrum was reduced with the UVES pipeline \citep{dekker-uves} and
exposures were combined and continuum fit with {\sc uves$\_$popler}
\citep{uvespopler, uvessquad}. During this process, the wavelengths
were vacuum and heliocentric velocity corrected. The spectrum covers
multiple lines for the $z_{\rm abs}=2.0708$ absorber such as {\Lya},
{\MgII}, and {\CIV}, which are further described in
Section~\ref{sec:absorption}.

With these data, the absorption system was characterized using the
methods presented in \citet{pointon19iso}. In short, absorption
profiles were modeled with {\sc vpfit} \citep{vpfit} to obtain
equivalent widths, column densities, and the kinematic
structure. Upper limits on equivalent width and column density
($3\sigma$) were measured by assuming a single cloud with a Doppler
parameter of $b\sim 8$~{\kms}. As shown in
Section~\ref{sec:absorption}, the upper limits on absorption are
located on the linear or near-linear part of the curve-of-growth, thus
the equivalent width limit is not sensitive to the choice of Doppler
parameter. The absorption redshift, $z_{\rm abs}$, is defined as the
optical depth-weighted median of absorption for {\MgII}.

\begin{figure*}
  \centering
  \includegraphics[width=\linewidth]{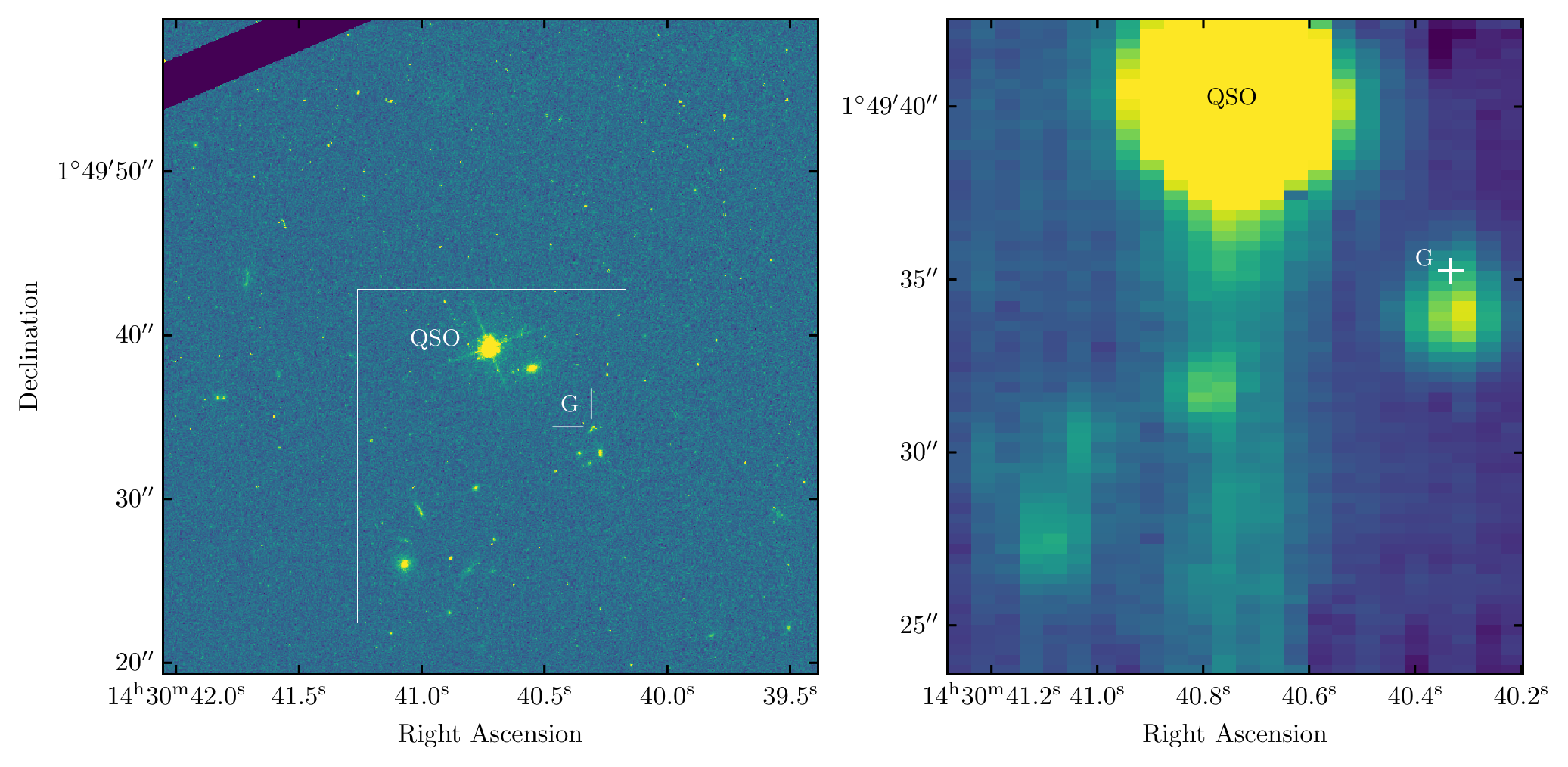}
  \caption{Overview of the quasar field J1430$+$014. (left) {\it
      HST}/ACS F625W image centered on the quasar. The KCWI field of
    view with the medium IFU slicer is marked with the white
    rectangle. (right) KCWI whitelight image, where the spectral
    direction is summed for each spaxel. The quasar (QSO) is the
    brightest object in both panels and the host galaxy (G) is marked
    by the cross. North is up and East is to the left in all images.}
  \label{fig:whitelight}
\end{figure*}

The total measured column densities, measured by summing the column
densities of the fitted Voigt profile (VP) components, were then
compared to the predicted column densities generated by {\sc Cloudy}
\citep{cloudy} to infer the total line-of-sight CGM metallicity. {\sc
  Cloudy} generates column density predictions given {\HI} column
densities ($N_{\tiny {\HI}}$), hydrogen densities ($n_{\rm H}$), and
metallicities ([Si/H]) by modeling a uniform layer of gas that is
irradiated by background UV radiation. Here we use the updated
\citet{haardt01} UV background (HM05 in {\sc Cloudy}) for comparison
to low redshift work. The more recent HM12 UV background
\citep{haardt12} is a harder spectrum, resulting in higher metallicity
estimates with a dependence on {\HI} column density at low redshift
\citep{chenUVB, wotta19, zahedy19LRG}. While the results using HM12
are also reported here for completeness, we focus on the HM05 results
for direct comparison to the large body of CGM metallicity work using
HM05 and to prevent any systematic biases that may be present when
comparing the results of the different UV backgrounds. We assumed a
single-phase model with no dust and with a solar abundance pattern
\citep{crighton13, crighton15, crighton16}. This single-phase model
forces the assumption that the majority of {\HI} is associated with
the low and intermediate ionization ions and this assumption is widely
used \citep[e.g.,][]{cooper15, fumagalli16, lehner16, lehner19,
  wotta16, wotta19, prochaska17, pointon19iso, pointon20}. Studying
the multiphase and multi-component nature of CGM absorbers suggests
that this assumption is valid, where \citet{muzahid15} found that
{\HI} is predominantly associated with the low ionization phase, with
small contributions from the high ionization phase. We then
constructed a likelihood function using the measured column densities
including upper limits and {\sc Cloudy} grids. Priors on the column
densities were applied and we used a Markov chain Monte Carlo (MCMC)
analysis with Bayesian statistics following the \citet{crighton15}
methods with \verb|emcee| \citep{mcmc} to generate posterior
distributions for $N_{\tiny {\HI}}$, $n_{\rm H}$, $U$, and
[Si/H]. Metallicities are reported as [Si/H] for comparison to low
redshift. Further details are presented in
Section~\ref{sec:metallicity}.

\subsection{{\it HST} Imaging and Galaxy Morphologies}
\label{sec:hstimage}

The field was imaged with {\it HST} using the F625W filter on the ACS
for 700~s (PID: 10576) and the data were reduced using the DrizzlePac
software \citep{drizzlepac}. We used the Source Extractor software
\citep[SExtractor;][]{bertin96} to measure the galaxy photometry with
a detection criterion of $1.5\sigma$ above the background. The
resulting magnitude is quoted in the AB system. The {\it HST} image
centered on J1430$+$014 is shown in the left panel of
Figure~\ref{fig:whitelight}.

Galaxy morphological parameters and orientations are modeled from the
{\it HST} image following the methods of \citet{kacprzak15}. In
summary, point spread functions (PSFs) for ACS images depend on both
time and position on the chip and the images also contain significant
geometrical distortions. Thus PSFs generated using Tiny Tim
\citep{tinytim} are appropriate for each galaxy in the field and are
then used in the galaxy modeling. Galaxy morphological parameters were
quantified by fitting a two-component disk$+$bulge model using {\sc
  GIM2D} \citep{simard02}. A disk component has an exponential profile
while the bulge has a S{\'e}rsic profile with $0.2\leq n\leq 4.0$. In
order to quantify the location of the line-of-sight through the CGM
relative to the modeled galaxy's on-the-sky orientation, we adopt the
standard convention that an azimuthal angle of $\Phi=0^{\circ}$ is
defined as the background quasar is located along the galaxy projected
major axis, while $\Phi=90^{\circ}$ is along the galaxy projected
minor axis. Additionally, a face-on galaxy is defined as having an
inclination of $i=0^{\circ}$, while an edge-on galaxy has
$i=90^{\circ}$. Further details are discussed in
Section~\ref{sec:galaxy}.

\subsection{KCWI Integral Field Spectroscopy}

Keck/KCWI observations of the J1430$+$014 field were conducted on 2018
February 15 UT (PID: 2018A\_W185) with the medium image slicer and BL
grating using a central wavelength of 4500~{\AA} and $2\times2$
binning. The medium slicer has a field of view (FOV) of
$16\farcs5\times20\farcs4$, resulting in a spatial sampling of
$0\farcs29\times 0\farcs69$, corresponding to $2.4\times5.8$~kpc at
$z\sim2$. The BL grating has a spectral resolution of $R\approx1800$
($\sim0.625$~{\AA} pix$^{-1}$) and spans $3500 \lesssim \lambda
\lesssim 5500$~{\AA} for our central wavelength setting. Four
exposures of 1300~s each (1.4~hrs total) were obtained on a single
pointing with a position angle of $0^{\circ}$. The KCWI footprint is
shown as the white rectangle on the {\it HST}/ACS image in the left
panel of Figure~\ref{fig:whitelight}. A single pointing was used in
this field to maximize the amount of time spent on galaxies observable
in the {\it HST}/ACS image. If a galaxy is not seen in the ACS image,
then its morphology cannot be measured, especially since the KCWI
spaxel sizes are on the order of $z\sim2$ galaxy sizes
\citep[half-light radii of $\sim2.5-3$~kpc;][]{allen17}. We have tiled
around the quasar in the rest of our fields to obtain a better census
of host galaxies.

The data were reduced using the publicly available KCWI Data Reduction
Pipeline\footnote{\dataset[https://github.com/Keck-DataReductionPipelines/KcwiDRP]{https://github.com/Keck-DataReductionPipelines/KcwiDRP}}
using default settings. Since separate sky fields were not obtained,
we masked the quasar, continuum objects, and any bright emission lines
for the automated sky subtraction step so that the sky estimate is not
skewed. Importantly, if the quasar is not removed before this step,
then its spectrum is over-subtracted from the datacube resulting in
false emission lines at wavelengths corresponding to strong
intervening absorption observed in the quasar spectrum. The data were
also flux-calibrated with standard star g191b2b from the KCWI DRP
starlist. Finally, wavelengths were vacuum and heliocentric velocity
corrected for direct comparison to the UVES quasar spectrum and the
four exposures were combined. The final datacube has a $3\sigma$ flux
limit of $3\times10^{-19}$~erg~s$^{-1}$~cm$^{-2}$~{\AA}$^{-1}$.

\begin{figure*}[ht]
  \includegraphics[width=\linewidth]{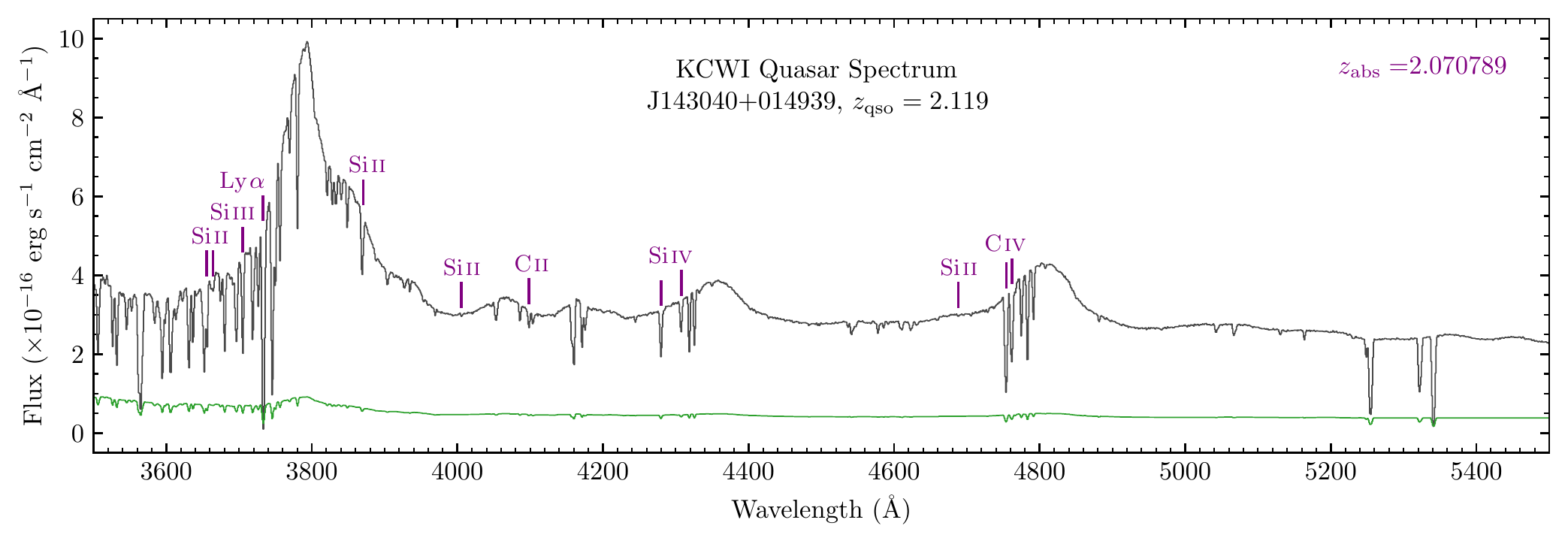}
  \caption{KCWI quasar spectrum with the BL grating for J1430$+$014
    summed over 32 spaxels. The data are plotted as the black
    histogram and the associated uncertainty (multiplied by 20) is
    plotted as the green line. Purple ticks and labels indicate the
    detected absorption features associated with the $z_{\rm
      abs}=2.0708$ absorption system. Further ions such as {\MgII} and
    {\FeII} are covered in the higher resolution VLT/UVES spectrum and
    all ions used in the metallicity analysis are plotted in
    Figure~\ref{fig:abssystem}.}
  \label{fig:kcwiqsospec}
\end{figure*}

\begin{figure*}[ht]
  \centering
  \includegraphics[scale=0.86]{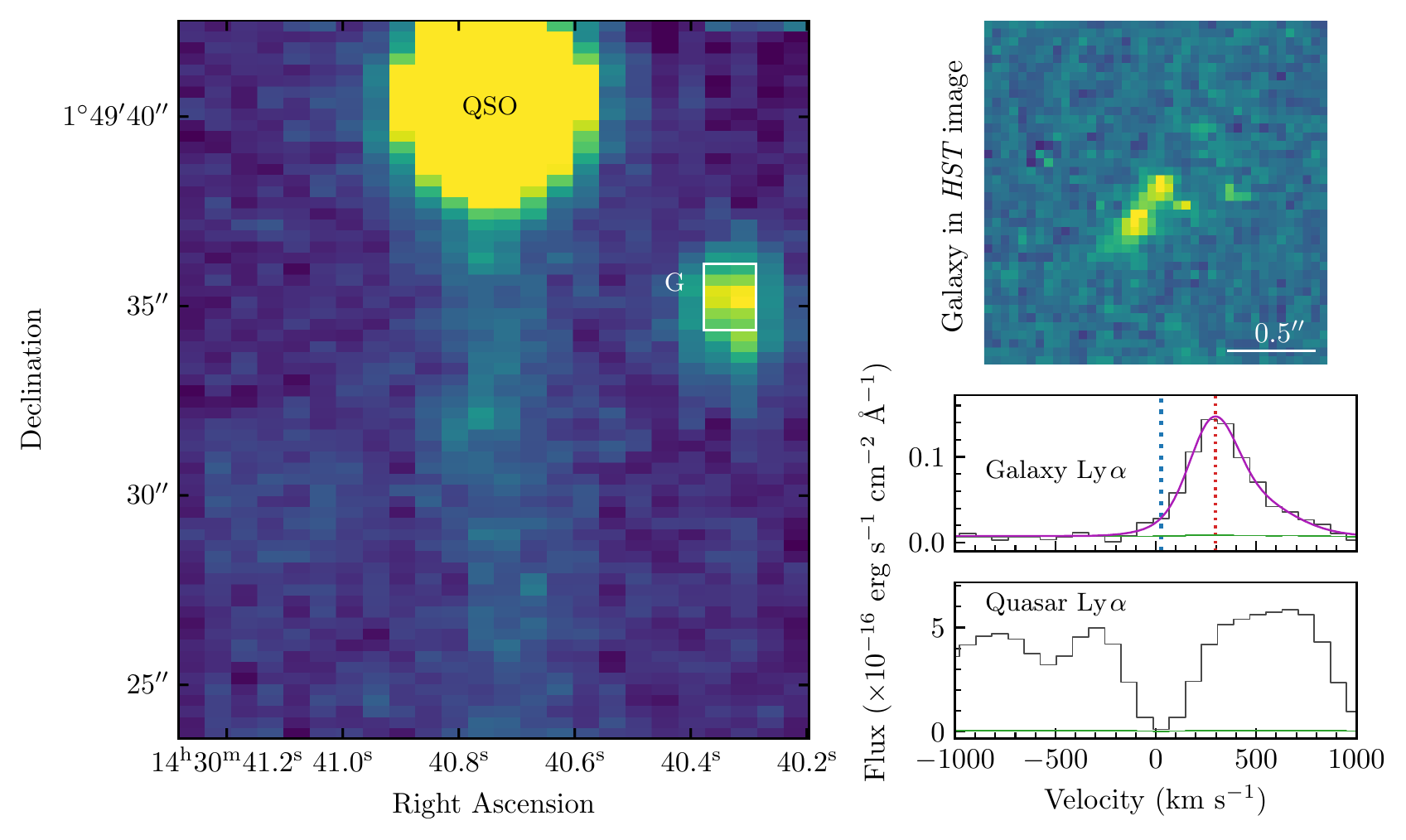}
  \caption{The galaxy associated with the observed absorption is
    located at $D=66$~kpc from the quasar sightline at $z_{\rm
      gal}=2.0711$. (left) KCWI slice within $v\pm1000$~{\kms} of the
    {\Lya} absorption. The quasar (QSO) is the bright object at the
    top of the field of view, while the host galaxy (G) is the bright
    region $7\farcs9$ away to the right side of the field of
    view. (upper right) The galaxy in the {\it HST} image, which is
    best modeled with an edge-on inclination
    ($i={85^{\circ}}^{+5}_{-2}$). The quasar sightline is oriented
    along the projected minor axis of the galaxy
    ($\Phi={89^{\circ}}^{+1}_{-5}$). (middle right) The galaxy {\Lya}
    emission line in the KCWI datacube summed over 12 spaxels (white
    box in the left panel). A double Gaussian fit is plotted as the
    purple curve. The wavelength at which the line peaks in flux
    measured from the Gaussian fit is plotted as the vertical red
    densely-dotted line while the final adopted galaxy redshift,
    $z_{\rm gal}$, is plotted as the vertical blue loosely-dotted
    line. (bottom right) {\Lya} absorption in the KCWI quasar spectrum
    over 32 spaxels for reference. The velocity zero point corresponds
    to the $z_{\rm abs}$ measured from {\MgII} absorption in the
    VLT/UVES quasar spectrum (see Figure~\ref{fig:abssystem}).}
  \label{fig:galaxy}
\end{figure*}

The resulting KCWI whitelight image, where the wavelength axis has
been collapsed, is shown in the right panel of
Figure~\ref{fig:whitelight}. The quasar is the brightest object in the
field and several continuum objects are observed corresponding to
galaxies in the {\it HST} image. Figure~\ref{fig:kcwiqsospec} presents
the quasar spectrum extracted from the KCWI datacube over 32
spaxels. The black and green lines represent the data and error
spectrum, respectively. The error spectrum has been multiplied by 20
in order to better show the variation across the spectrum. The purple
annotations indicate absorption features associated with the $z_{\rm
  abs}=2.0708$ {\MgII} and {\CIV} absorbers found in the VLT/UVES
spectrum. Additionally, {\Lya} absorption is clearly observed in the
KCWI quasar spectrum, providing a reference for host galaxy {\Lya}
emission features with similar resolution.

\section{Results}
\label{sec:results}

In this section, we present the first results of our CGM at Cosmic
Noon with KCWI program for J$1430+014$, $z_{\rm abs}=2.0708$. In this
field, we search for and find the absorbing host galaxy, characterize
the galaxy's photometric properties, measure the absorption
properties, estimate the CGM metallicity, and model the observed gas
as outflowing material.

\subsection{Host Galaxy}
\label{sec:galaxy}

To find the absorber host galaxy, we searched a narrow band of
$v=\pm1000$~{\kms} around the absorber redshift for {\Lya}
emission. Figure~\ref{fig:galaxy} presents the results of this search,
where the left panel shows the KCWI narrow band image. The quasar is
the brightest source in the field and the host galaxy {\Lya} emission
is located $\theta=7\farcs9$ southwest of the quasar. The bright knot
of emission corresponds to the galaxy from the {\it HST}/ACS image
shown in the top right panel. No obvious emission from other galaxies
in the {\it HST} image is observed in this field out to $D<150$~kpc.

\begin{deluxetable}{lcl}
  \tablecaption{Host Galaxy Properties \label{tab:galaxy}}
  \tablehead{
    \colhead{Property} &
    \colhead{Value}    &
    \colhead{Units}
  }
  \startdata
  RA                 & \phs14:30:40.30     &                     \\
  Dec                & $+$01:49:34.33      &                     \\
  $z_{\rm gal}$\tablenotemark{a} & $2.0711\pm0.0008$ &             \\
  $D$                & $66.4\pm0.3$        & kpc                 \\
  $m_{\rm F625W}$      & $24.0\pm0.4$        & AB mag               \\
  $M_B$              & $-22.7_{-0.8}^{+0.7}$  &                      \\
  $L_B/L_B^{\ast}$     & $1.7_{-0.9}^{+1.9}$   &                      \\
  $W_{r}(\Lya)$       & $44.4$               & {\AA}               \\
  $L_{\tiny {\Lya}}$    & $2.32\times10^{42}$  & erg~s$^{-1}$         \\
  ${\rm SFR}_{\rm FUV}$ & $37.8$              & M$_{\odot}$~yr$^{-1}$  \\
  ${\rm SFR}_{\tiny {\Lya}}$\tablenotemark{b} & $>8.4$ & M$_{\odot}$~yr$^{-1}$ \\
  $i$\tablenotemark{c} & $85^{+5}_{-2}$      & degrees              \\
  $\Phi$             & $89^{+1}_{-5}$        & degrees              \\[2pt]
  \enddata

  \tablenotetext{a}{We shifted the galaxy redshift blueward by $v_{\rm
      red}=273$~{\kms} from $\lambda_{\rm obs}=3736.79$~{\AA} due to
    resonant scattering using: $z_{\rm gal} = z_{\tiny \Lya} - (v_{\rm
      red}/c)(\lambda_{\rm obs}/1215.67)$.}

  \tablenotetext{b}{Assuming $f_{\rm esc, LyC}=0.0$ as a conservative
    lower limit. If we assume $f_{\rm esc, LyC}=0.15$,
    we measure ${\rm SFR}_{\tiny {\Lya}}=15.8$~M$_{\odot}$~yr$^{-1}$.}

  \tablenotetext{c}{The galaxy is assumed to be a single clumpy galaxy
    rather than a major merger. See Section~\ref{sec:galmorph} for
    further discussion.}

\end{deluxetable}

\begin{figure*}[t]
  \includegraphics[width=\linewidth]{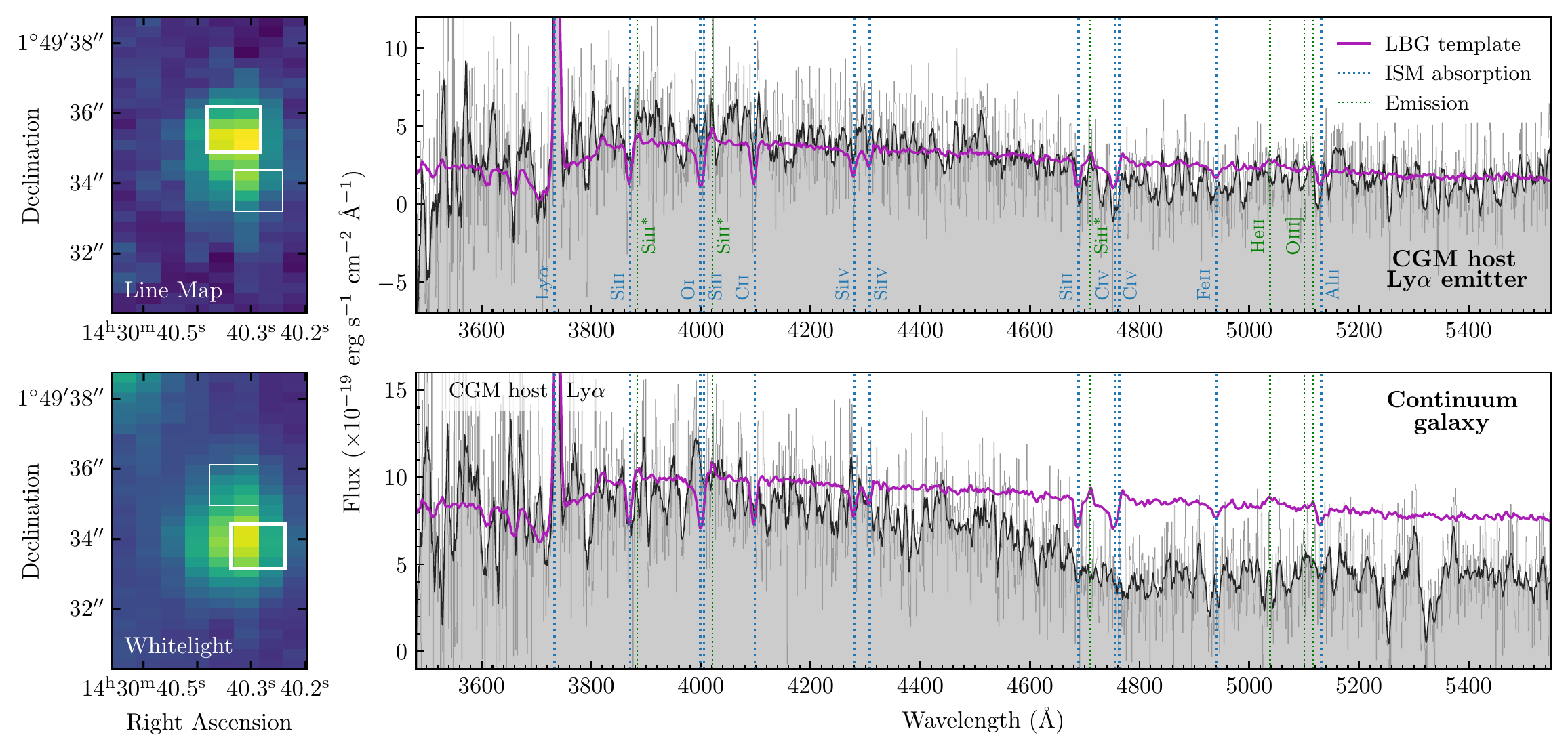}
  \caption{Comparison between the {\Lya} emitting host galaxy (top)
    and an unrelated bright continuum galaxy contaminated with {\Lya}
    emission (bottom). Left panels are a zoom-in view of the KCWI
    datacube, where the top image is a line map centered on
    $v\pm1000$~{\kms} around $z_{\rm abs}$ (see
    Figure~\ref{fig:galaxy}). The bottom slice shows the bright
    continuum source which is physically offset from the {\Lya}
    emission by $\sim1\farcs6$ (see
    Figure~\ref{fig:whitelight}). Thick white boxes indicate the
    spaxels used to create the corresponding spectra on the right,
    while the thin white boxes are plotted for comparison. The spaxels
    chosen are offset slightly to reduce the amount of
    cross-contamination in the spectra. (right) The thin black lines
    are the data summed over eight spaxels in the KCWI datacube and
    the shaded region is plotted for ease in identifying absorption
    features. Both spectra have been smoothed using a boxcar filter
    over eight spectral pixels (thick black line). The Lyman break
    galaxy template from \citet{shapley03} is plotted in purple, where
    the wavelengths are shifted according to $z_{\rm gal}=2.0711$ and
    the flux is adjusted to roughly match the flux surrounding the
    {\Lya} emission line. Vertical dotted lines indicate strong ISM
    absorption and emission lines in the LBG template. The CGM host
    galaxy is well-fitted by a LBG template, whereas the continuum
    galaxy lacks the ISM absorption features required to classify as a
    $z\sim2$ LBG and the low continuum level at higher wavelengths
    indicates this is a lower redshift object. Thus the {\Lya}
    emission in the bottom panel is likely contamination from the CGM
    host galaxy.}
  \label{fig:galspec}
\end{figure*}

The galaxy {\Lya} emission and CGM {\Lya} absorption from the quasar
spectrum, both of which are 1D extracted spectra from the KCWI cube,
are presented in the bottom right panels of Figure~\ref{fig:galaxy}
(see Figure~\ref{fig:galspec} for the full galaxy spectrum). The
velocity zero point corresponds to the optical depth-weighted median
of {\MgII} absorption in the UVES spectrum. We selected 12 spaxels
around the galaxy for extraction, which roughly corresponds to the
full width of the seeing (FWHM$\sim1\arcsec$), and these spaxels are
indicated by the white box in the left panel of
Figure~\ref{fig:galaxy}. Because {\Lya} emission is subject to
resonant scattering and thus the profile depends on the amount and
kinematics of gas the photons must pass through, the observed redshift
is not necessarily the systemic galaxy redshift
\citep[e.g.,][]{zheng02, shapley03, rakic11, trainor15}. Therefore, we
use the Method 2 red peak relation described in \citet{verhamme18},
$v_{\rm red}=0.8\times{\rm FWHM}({\Lya})-34$~{\kms}, to calculate the
corrected galaxy redshift. We first obtain the observed {\Lya}
emission wavelength by fitting a double Gaussian profile (purple line
in the middle right panel of Figure~\ref{fig:galaxy}) and finding the
wavelength corresponding to the max flux of the line, $\lambda_{\rm
  obs}=3736.79$~{\AA}. This wavelength is indicated in the panel by a
vertical red dotted line. The full width at half maximum of this
galaxy's {\Lya} emission line is then ${\rm FWHM}({\Lya})=341$~{\kms},
which results in a redshifted velocity offset of $v_{\rm
  red}=273$~{\kms} from the actual systemic galaxy redshift. This
velocity was then applied with: $z_{\rm gal} = z_{\tiny \Lya} -
(v_{\rm red}/c)(\lambda_{\rm obs}/1215.67)$. With this correction, we
find that the host galaxy is located at $z_{\rm gal}=2.0711\pm0.0008$
and this is plotted as the vertical blue dotted line in the middle
right panel of Figure~\ref{fig:galaxy}.

Full host galaxy properties are tabulated in
Table~\ref{tab:galaxy}. The galaxy is located at an impact parameter
of $D=66$~kpc from the quasar sightline, which is reasonable compared
to low redshift {\MgII} samples. We converted the apparent magnitude,
$m_{\rm F625W}$, to an absolute $B$-band magnitude with a
$k$-correction that assumes an Scd galaxy spectral energy distribution
\citep[e.g.,][]{magiicat1} and report uncertainties corresponding to
values derived using Sbc and Im galaxy SEDs in the Table. The absolute
magnitude is then $M_B=-22.7$, corresponding to $L_B/L_B^{\ast}=1.7$
using the relation between $M^{\ast}$ and redshift from
\citet{gabasch04}, which measures $B$-band luminosity functions over
the range $0.5<z<5.0$.

From an inspection of the {\it HST} image, it is clear that the galaxy
has a clumpy morphology, with a bright region on both sides of a
disk-like structure. Galaxies at $z>1$ typically have clumpy
morphologies due to large star-forming regions observed at rest-frame
UV wavelengths (i.e., the F625W band in Figure~\ref{fig:whitelight})
and these irregular morphologies may be indistinguishable from
mergers. The galaxy is similar to ``double'' or ``chain'' clump
morphologies found in deep ($\sim$100,000~s) {\it HST}/ACS rest-frame
UV images in the Hubble Ultra Deep Field
\citep[e.g.,][]{elmegreen07}. Because of the short (700~s) exposure in
the F625W band (rest-frame UV) of the J1430$+$014 field, it is not
immediately clear if this galaxy is a clumpy edge-on disk or two small
spheroidal galaxies undergoing a major merger. The major merger
scenario appears to be less likely for the following reasons: 1) The
object size is on the order of typical $z=2$ star-forming galaxy sizes
\citep[$\sim 2.5-3$~kpc;][]{vanderwel14, allen17} but the clumps have
a smaller separation, suggesting that a single object is more
likely. 2) Quiescent galaxies are on average smaller, with sizes that
are more consistent with the clump sizes in this object
\citep{vanderwel14}, but the object is star-forming as indicated by
the presence of {\Lya} and the discussion later in this section. Thus
the quiescent galaxy sizes are not relevant here. 3) The probability
of a galaxy at $z=2$ having a companion within a projected separation
of $0\farcs23$ or 1.9~kpc, which is the separation of the two clumps
in our object, is $0-3\%$ for galaxies in the Illustris simulations
\citep{simons19}. Our full rationale for a single, clumpy galaxy
assumption is discussed in Section~\ref{sec:galmorph}. For the rest of
our analysis, we assume the object is a single, clumpy galaxy that is
not undergoing a major merger.

From the GIM2D modeling described in Section~\ref{sec:hstimage}, we
find that this object, assuming it is a single clumpy galaxy, is best
modeled as an edge-on galaxy with $i=85^{\circ}$. The modeling also
indicates that the quasar sightline probes the projected minor axis of
this galaxy with $\Phi=89^{\circ}$. Observations and simulations
suggest that outflowing gas is primarily observed as bipolar cones
aligned with the minor axis of galaxies
\citep[e.g.,][]{outflowsreview, bordoloi11, bouche12, kcn12,
  kacprzak14, martin12, martin19, rubin-winds, rubin-winds14, shen13,
  fox15, schroetter16, nelson19}. Thus, assuming an edge-on
morphology, this galaxy presents an ideal case for studying outflows
in the CGM at peak star formation activity.

We measure a star formation rate (SFR) using two methods. The first
method uses the relation between {\Lya} luminosity and equivalent
width presented in \citet{sobral19}, who showed that ${\rm SFR}_{\tiny
  \Lya}$ can be measured from this relation over a wide redshift range
($0<z<2.6$) and is consistent with dust-corrected {\Ha} SFRs. This
method assumes a Salpeter IMF and a case B recombination ratio between
{\Lya} and {\Ha}, but we convert to a Kroupa IMF by multiplying by
0.62 \citep[e.g.,][]{speagle14}. We obtain ${\rm SFR}_{\tiny \Lya} =
8.4\pm0.8$~M$_{\odot}$~yr$^{-1}$ by assuming $f_{\rm esc,LyC}=0.0$,
the escape fraction of Lyman continuum photons. The value used for
$f_{\rm esc,LyC}$ does affect the SFR, with larger escape fractions
resulting in larger SFRs, but a value of 0.0 is typically assumed
\citep{sobral19}, providing a conservative lower limit. If we use
$f_{\rm esc,LyC}=0.15$, then ${\rm SFR}_{\tiny
  \Lya}=15.8\pm1.3$~M$_{\odot}$~yr$^{-1}$ \citep[for further
  discussion of the value for {\Lya} emitters, see][]{matthee17,
  verhamme17}.

The second method calculates an unobscured SFR using the far-UV (FUV)
continuum. We used the method of \citet{hao11}, assuming a Kroupa IMF,
solar metallicity, and 100 Myr age. We estimated the dust attenuation
by measuring a NUV flux from the {\it HST}/ACS image and the FUV flux
from the KCWI datacube. From this, we obtain ${\rm SFR}_{\rm
  FUV}=37.8$~M$_{\odot}$~yr$^{-1}$. Because of the uncertainty in the
escape fraction of LyC photons, we adopt this SFR for the
galaxy. Furthermore, the galaxy disk is parallel to our line-of-sight
where dust blocks much of the light, which likely means that both the
{\Lya} and the FUV luminosities are underestimated even despite the
dust correction. This is corroborated in the {\sc Simba} simulations,
where the average $z\sim2$ star-forming galaxy with an absolute
$B$-band magnitude comparable to the J1430$+$014, $z_{\rm gal}=2.0711$
galaxy has ${\rm SFR}\sim85$~M$_{\odot}$~yr$^{-1}$. The SFR measured here is
most likely a lower limit.

From further examination of the {\it HST} image (left panel of
Figure~\ref{fig:whitelight}), it appears possible that this galaxy is
located in a group environment with three galaxies just $\sim1\farcs8$
southward, particularly the galaxy that is bright in the KCWI
whitelight image. Figure~\ref{fig:galspec} explores the full spectra
of both the identified host galaxy (top panels) and the bright
continuum galaxy (bottom panels). The spectra are summed over eight
spaxels indicated by the white boxes on the KCWI image zoom-ins (left)
and are smoothed using a boxcar filter over eight voxels (spectral
pixels). The spaxels are chosen to select a majority of the light from
each galaxy but still be separated from each other enough to reduce
potential cross-contamination due to the seeing and large {\Lya}
halos. If we assume that both galaxies are located at $z_{\rm
  gal}=2.0711$ and both contribute to the observed {\Lya} emission,
then we can overlay the template spectrum for a Lyman break galaxy
(LBG) from \citet{shapley03}. This method for determining galaxy
redshifts in low resolution data is often used for high-redshift
objects \citep[e.g.,][]{cooke05, cooke06}. The CGM host galaxy (top)
matches the LBG template spectrum continuum level. Strong ISM
absorption features such as {\OI}, {\CII}, {\SiII}, {\SiIV}, and
{\CIV}, and the {\SiII}$^{\ast}$ ISM emission lines appear to be
present, though the spectrum is noisy. These all indicate that the
measured redshift is likely real. In contrast, the continuum galaxy
(bottom) does not match the continuum level across the wavelength
range, with the lower flux at higher wavelength indicating that this
is likely a lower redshift object. Nor does the continuum galaxy have
the ISM features present, particularly {\CIV} absorption. This is true
for all three galaxies south of the host galaxy in the HST image. We
therefore conclude that the host galaxy has a large {\Lya} halo
(either physically or due to the seeing) out to roughly $1\farcs8$ and
that it is likely isolated to the limits of our data. We assume the
galaxy is isolated for the rest of our analysis.

\begin{figure*}[ht]
  \includegraphics[width=\linewidth]{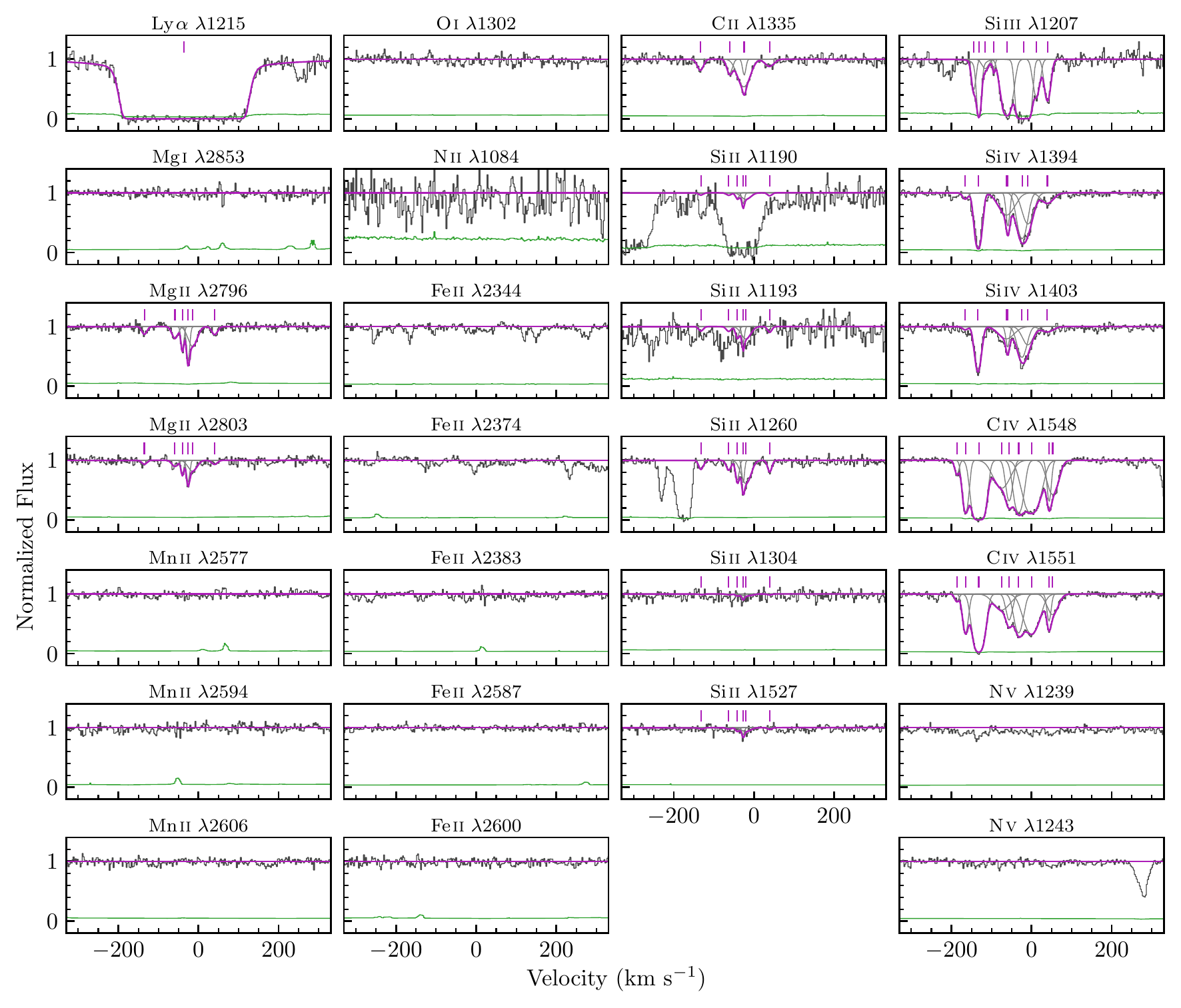}
  \caption{Absorption system at $z_{\rm abs}=2.070789$ in the VLT/UVES
    spectrum of J143040$+$014939 covering the ions used in the
    metallicity analysis. The data for the ion transitions labeled
    above each panel are plotted as black histograms and the error
    spectra are the green histograms. The velocity zero point is
    $z_{\rm gal}=2.0711$. Voigt profile fits are plotted as purple
    lines while the purple ticks and gray lines indicate the
    individual Voigt profile components. Transitions in which only an
    upper limit on absorption could be measured have a purple horizontal
    line plotted through the data with no associated tick marks. With
    the exception of {\Lya}, the strongest absorption is found in
    higher ions such as {\SiIII}, {\SiIV}, and {\CIV}.}
  \label{fig:abssystem}
\end{figure*}

\subsection{Absorption System}
\label{sec:absorption}

From the VLT/UVES quasar spectrum, we identified several absorption
lines associated with this host galaxy. The absorption system is
plotted in Figure~\ref{fig:abssystem}, where the UVES data and
uncertainties are plotted as the black and green lines,
respectively. While the quasar spectrum covers {\Lya}, which is
important for metallicity analyses, neither the UVES nor KCWI spectra
cover the rest of the Lyman series lines. Typical CGM metal lines such
as {\MgII}, {\CII}, {\SiII}, {\SiIII}, {\SiIV}, and {\CIV} are
detected and we further have coverage of, but no detected absorption
for, {\MgI}, {\OI}, {\MnII}, {\NII}, {\FeII}, and {\NV}. Each detected
unsaturated line has similar kinematics, with weaker components on
either side of the bulk of the absorption near $z_{\rm gal}$, but the
ionization structure varies across the profile where the blueshifted
weaker components are relatively stronger in {\SiIII} and {\SiIV}
compared to {\SiII}. These line-of-sight structure variations are
further discussed in Section~\ref{sec:metaldisc}, but for comparison
to the majority of the CGM metallicity literature, we focus on a total
metallicity, which averages out the structure variations. As stated in
Section~\ref{sec:qsospec}, we model all absorption using {\sc vpfit}
and describe our preferred fitting philosophy here.

We first fit the metal lines, where we took the approach of fitting
the minimum number of VP components to obtain a reasonable chi-squared
value \citep[see e.g.,][]{cvc03, churchill20, evans-thesis}. We
assumed that doublets and multiplets for a given ion have the same
kinematic structure (i.e., the {\MgIIdblt} doublet), but we did not
require similar kinematic structure across ions (i.e., {\MgII}
compared to {\SiII} or {\CII}). The approach requiring similar
kinematic structure across ions has been applied by the KBSS for
understanding the thermal state of the $z=2-3$ CGM \citep{rudie19},
but we focus instead on obtaining the most accurate total column
density for each ion.

Due to the fact that {\Lya} is saturated and the rest of the Lyman
series lines are not covered in the spectrum, we constrained the {\HI}
column density using a combination of two methods \citep[also
  see][]{pointon19iso}. We first fit a single VP component to {\Lya}
and adopted this value as an upper limit. The lack of damping wings on
the profile indicates that this absorber is not a DLA or sub-DLA and
the best-fitting column density, $\log N_{\tiny {\HI}}=18.18$~{\cmsq},
indicates that this is most likely a LLS. We employed a second method
where we applied the fit to {\MgII} as a template and fixed the VP
component redshifts. The absorption system was then assumed to be
dominated by thermal broadening, such that the ratio between the
Doppler $b$ parameters of {\Lya} and {\MgII} for a given VP component
is defined by the ratio between the atomic masses. This resulted in a
value of $\log N_{\tiny {\HI}}=16.37$~{\cmsq}. The {\HI} column
density is likely between the measurements with these two methods, but
it is difficult to determine confidently due to the lack of additional
Lyman series lines. To be conservative in our {\HI} column density
range, we therefore adopted a lower limit of $\log N_{\tiny
  {\HI}}=15.00$~{\cmsq}.

\begin{deluxetable}{lcrr}
  \tablecaption{Rest Absorption Equivalent Widths\label{tab:ews}}
  \tablehead{
    \colhead{Transition} &
    \colhead{EW} &
    \colhead{$v_-$\tablenotemark{a}} &
    \colhead{$v_+$\tablenotemark{a}} \\[-2pt]
    \colhead{} &
    \colhead{({\AA})} &
    \colhead{(km~s$^{-1}$)} &
    \colhead{(km~s$^{-1}$)} 
  }
  \startdata
      {\Lya}   $\lambda1215$ & $1.472\pm0.013$ & $-514$   & $+381$   \\
      {\MgI}   $\lambda2853$ & $<0.019$        & $\cdots$ & $\cdots$ \\
      {\OI}    $\lambda1302$ & $<0.009$        & $\cdots$ & $\cdots$ \\
      {\MgII}  $\lambda2796$ & $0.241\pm0.010$ & $-151$   & $ +56$   \\
      {\MgII}  $\lambda2803$ & $0.137\pm0.010$ & $-147$   & $ +52$   \\
      {\MnII}  $\lambda2577$ & $<0.010$        & $\cdots$ & $\cdots$ \\
      {\MnII}  $\lambda2594$ & $<0.011$        & $\cdots$ & $\cdots$ \\
      {\MnII}  $\lambda2606$ & $<0.011$        & $\cdots$ & $\cdots$ \\
      {\FeII}  $\lambda2344$ & $<0.008$        & $\cdots$ & $\cdots$ \\
      {\FeII}  $\lambda2374$ & $<0.009$        & $\cdots$ & $\cdots$ \\
      {\FeII}  $\lambda2383$ & $<0.008$        & $\cdots$ & $\cdots$ \\
      {\FeII}  $\lambda2587$ & $<0.009$        & $\cdots$ & $\cdots$ \\
      {\FeII}  $\lambda2600$ & $<0.011$        & $\cdots$ & $\cdots$ \\
      {\SiII}  $\lambda1190$ & $0.029\pm0.009$ & $-144$   & $ +46$   \\
      {\SiII}  $\lambda1193$ & $0.053\pm0.010$ & $-144$   & $ +48$   \\
      {\SiII}  $\lambda1260$ & $0.104\pm0.004$ & $-147$   & $ +50$   \\
      {\SiII}  $\lambda1304$ & $0.011\pm0.006$ & $-137$   & $ +43$   \\
      {\SiII}  $\lambda1527$ & $0.023\pm0.005$ & $-140$   & $ +47$   \\
      {\CII}   $\lambda1335$ & $0.136\pm0.005$ & $-153$   & $ +62$   \\
      {\NII}   $\lambda1084$ & $<0.034$        & $\cdots$ & $\cdots$ \\
      {\SiIII} $\lambda1207$ & $0.534\pm0.010$ & $-175$   & $ +60$   \\
      {\SiIV}  $\lambda1394$ & $0.405\pm0.005$ & $-181$   & $ +74$   \\
      {\SiIV}  $\lambda1403$ & $0.261\pm0.005$ & $-179$   & $ +69$   \\
      {\CIV}   $\lambda1548$ & $0.957\pm0.004$ & $-200$   & $ +89$   \\
      {\CIV}   $\lambda1551$ & $0.692\pm0.004$ & $-199$   & $ +86$   \\
      {\NV}    $\lambda1239$ & $<0.005$        & $\cdots$ & $\cdots$ \\
      {\NV}    $\lambda1243$ & $<0.006$        & $\cdots$ & $\cdots$ \\[2pt]
      \enddata
      
      \tablenotetext{a}{Velocity boundaries on absorption, defined as
        the velocity at which the VP model recovers to within $1\%$ of
        the continuum. The values are relative to $z_{\rm
          gal}=2.0711$. To compare to $z_{\rm abs}$, which is defined
        as the optical depth-weighted median of the
        {\MgII}~$\lambda2796$ line, the conversion is $v_{\rm
          abs}=v_{\tiny (+/-)}+27$~km~s$^{-1}$.}

\end{deluxetable}

\begin{deluxetable}{lcc}
  \tablecaption{Measured Column Densities\label{tab:columns}}
  \tablehead{
    \colhead{Ion} &
    \colhead{\hspace{2cm}\hspace{2cm}} &
    \colhead{$\log N$ (cm$^{-2}$)}
  }
  \startdata
      {\HI}~\tablenotemark{$a$} && $[15.00, 18.18]$ \\
      {\MgI}   && $<11.18$  \\
      {\OI}    && $<13.13$  \\
      {\MgII}  && $12.88\pm0.02$  \\
      {\MnII}  && $<11.70$  \\
      {\FeII}  && $<11.74$  \\
      {\SiII}  && $12.95\pm0.03$  \\
      {\CII}   && $13.92\pm0.01$  \\
      {\NII}   && $<13.62$  \\
      {\SiIII} && $14.85\pm0.63$  \\
      {\SiIV}  && $13.91\pm0.01$  \\
      {\CIV}   && $14.79\pm0.01$  \\
      {\NV}    && $<12.41$  \\[2pt]
  \enddata
      
  \tablenotetext{a}{{\Lya} is saturated and no other Lyman series
    lines are covered, so we defined boundaries on this value. See
    Section~\ref{sec:absorption} for details.}

\end{deluxetable}

The results of our VP fitting are shown in
Figure~\ref{fig:abssystem}. The purple line demonstrates the VP model
to the data, where individual VP components are plotted as gray curves
and their velocity centroids are purple ticks. The plotted {\Lya} fit
is derived from the single component fit described above. Where
absorption was not formally detected, we plot a horizontal purple line
with no VP components. The velocity zero point is $z_{\rm gal}$, which
is redshifted from the optical depth-weighted median of absorption for
the {\MgII}~$\lambda2796$ line, $z_{\rm abs}$, by $27$~{\kms}.

Equivalent widths and velocity bounds relative to $z_{\rm gal}$ for
each transition are tabulated in Table~\ref{tab:ews}. These values
were measured from the VP models instead of the data due to unrelated
absorption in the data for several transitions (i.e., {\SiII}
$\lambda1190, \lambda1193, \lambda1260$). Equivalent width upper
limits are reported at the $3\sigma$ level and were calculated using
the observed spectrum to account for noise. The boundaries on
absorption are defined as the velocity at which the VP model recovers
to within $1\%$ of the continuum. It is interesting to note that the
{\MgII} equivalent width is consistent with the $z<1$ $W_r(2796)-D$
anti-correlation and log-linear fit from \citet{magiicat2,
  magiicat1}. A larger sample of $z=2$ absorber--galaxy pairs will
investigate whether the relation evolves with redshift, where
\citet{chen12} did not find any evolution.

The measured column densities for each ion are listed in
Table~\ref{tab:columns}. The table lists two values for {\HI}, which
are the boundary values adopted from the fitting method described
above. Column density upper limits were calculated using the $3\sigma$
equivalent width upper limits and assuming a Doppler parameter of
$b=8$~{\kms}, which is the typical value for {\SiII} in the
\citet{pointon19iso} sample. The choice of Doppler parameter does not
change the measured value.

The kinematic structure of the metal lines is reminiscent of the
structure found in the Milky Way Fermi bubbles \citep{fox15,
  bordoloi17}, where there is strong absorption near the systemic
galaxy velocity and several higher velocity components that are
roughly symmetrically redshifted and blueshifted. Recall that the host
galaxy is assumed to be an edge-on galaxy probed along its minor axis,
which is similar to the Milky Way and Fermi bubble
geometry. Alternatively, the host galaxy could instead be a major
merger, with the projected major axis of the merger aligned
perpendicular to the quasar sightline. \citet[][]{rupkemakani} found
bipolar outflow bubbles with an hourglass shape that are consistent
with the structure of the Milky Way Fermi Bubbles in emission out to
50~kpc in a merger system \citep[see also][]{burchett20}. Within this
structure they found evidence for two outflow episodes that took place
0.4~Gyr ago and 7~Gyr ago. Given that the impact parameter of the
system we study here at 66~kpc is not much further than the Rupke et
al. system, the quasar sightline is perpendicular to the projected
major axis of the galaxy, the absorption roughly symmetrically spans
the galaxy redshift, and the galaxy is star-forming, it is possible
that the quasar sightline is probing an outflowing cone of gas, with
physical edges defined by the velocity edges of the absorption (see
Table~\ref{tab:ews}) regardless of the galaxy morphology. We further
explore the observed kinematics with an outflow model in
Section~\ref{sec:outflowmod} and other scenarios that might describe
the absorption properties in Section~\ref{sec:othergasflows}.

\begin{figure}[t]
  \includegraphics[width=\linewidth]{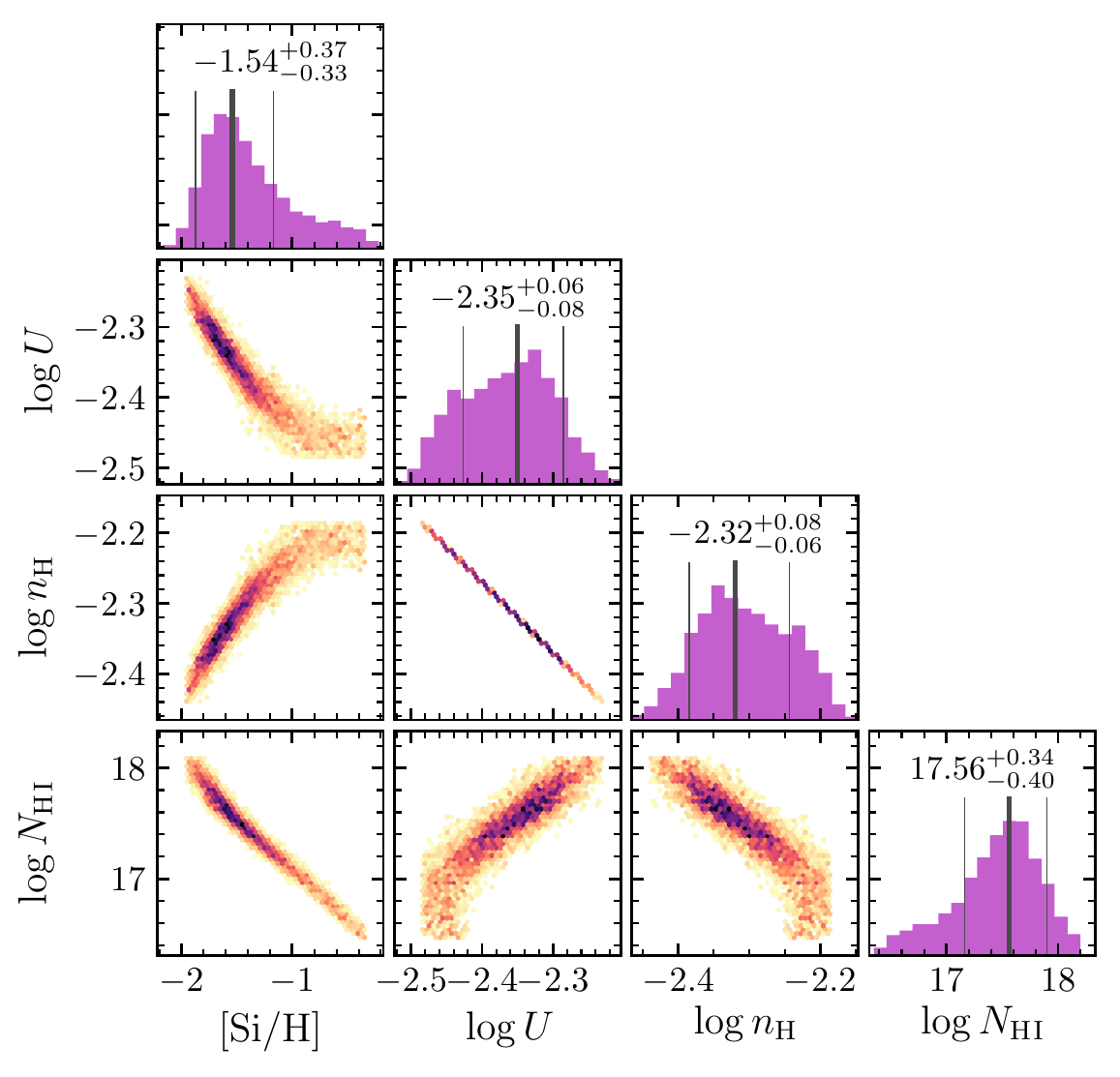}
  \caption{Posterior distributions from the MCMC modeling for
    metallicity, [Si/H], ionization parameter, $\log U$, hydrogen
    density, $\log n_{\rm H}$, and {\HI} column density, $\log
    N_{\tiny {\HI}}$ for the All Ions run with HM05. Histograms show
    the sample distribution for each quantity, where the thin vertical
    lines indicate the smallest interval containing 68\% of the MCMC
    samples and the thick vertical lines indicate the maximum
    likelihood value. We adopt a metallicity of $[{\rm Si}/{\rm
        H}]=-1.5^{+0.4}_{-0.3}$.}
  \label{fig:corner}
\end{figure}

\subsection{CGM Metallicity}
\label{sec:metallicity}

In order to infer a total line-of-sight CGM metallicity for this
absorber, we compared the observed column densities listed in
Table~\ref{tab:columns} to the predicted column densities generated by
{\sc Cloudy} with the HM05 UV background using an MCMC analysis and
Bayesian statistics (for further details, see
Section~\ref{sec:qsospec} and \citealp{pointon19iso, pointon20,
  crighton15, crighton16} as well as \citealp{cooper15, fumagalli16,
  wotta16, wotta19, prochaska17, lehner19}). The input {\sc Cloudy}
grids span ranges of $-5.0<\log n_{\rm H}<-2.0$~cm$^{-3}$, $13.0<\log
N_{\tiny \HI}<19.5$~{\cmsq}, and $-3.0<[{\rm Si}/{\rm H}]<1.0$ with
step sizes of 0.5 and are used as flat priors in the MCMC
analysis. After testing, a minimum value of $\log n_{\rm H}>-3.1$ was
enforced to avoid MCMC walkers getting stuck in local maxima (locally
higher likelihood or probability). Nearly all LLSs studied in
\citet{lehner16} were modeled to have hydrogen densities above this
minimum value, so this lower limit is reasonable. Priors on the
measured column density for each metal line are assumed to be a
Gaussian distribution defined by the measured column density and its
uncertainties as listed in Table~\ref{tab:columns}. In the case that
only an upper limit could be measured, these values were applied as
one-sided Gaussian priors with a sigma of 5. For {\HI}, the bounds on
$\log N_{\tiny \HI}$ were applied as bounds on a flat prior, which
allows the MCMC analysis to find a best-fit solution anywhere in that
range. We assume a single-phase model in which column density
measurements and upper limits for all ions listed in
Table~\ref{tab:columns} were used. We explore how removing the highest
ionization lines affects the inferred total metallicity of the low
ionization gas below and additionally report the inferred values using
the HM12 UV background for completeness in
Table~\ref{tab:metallicity}.

\begin{figure}[t]
  \includegraphics[width=\linewidth]{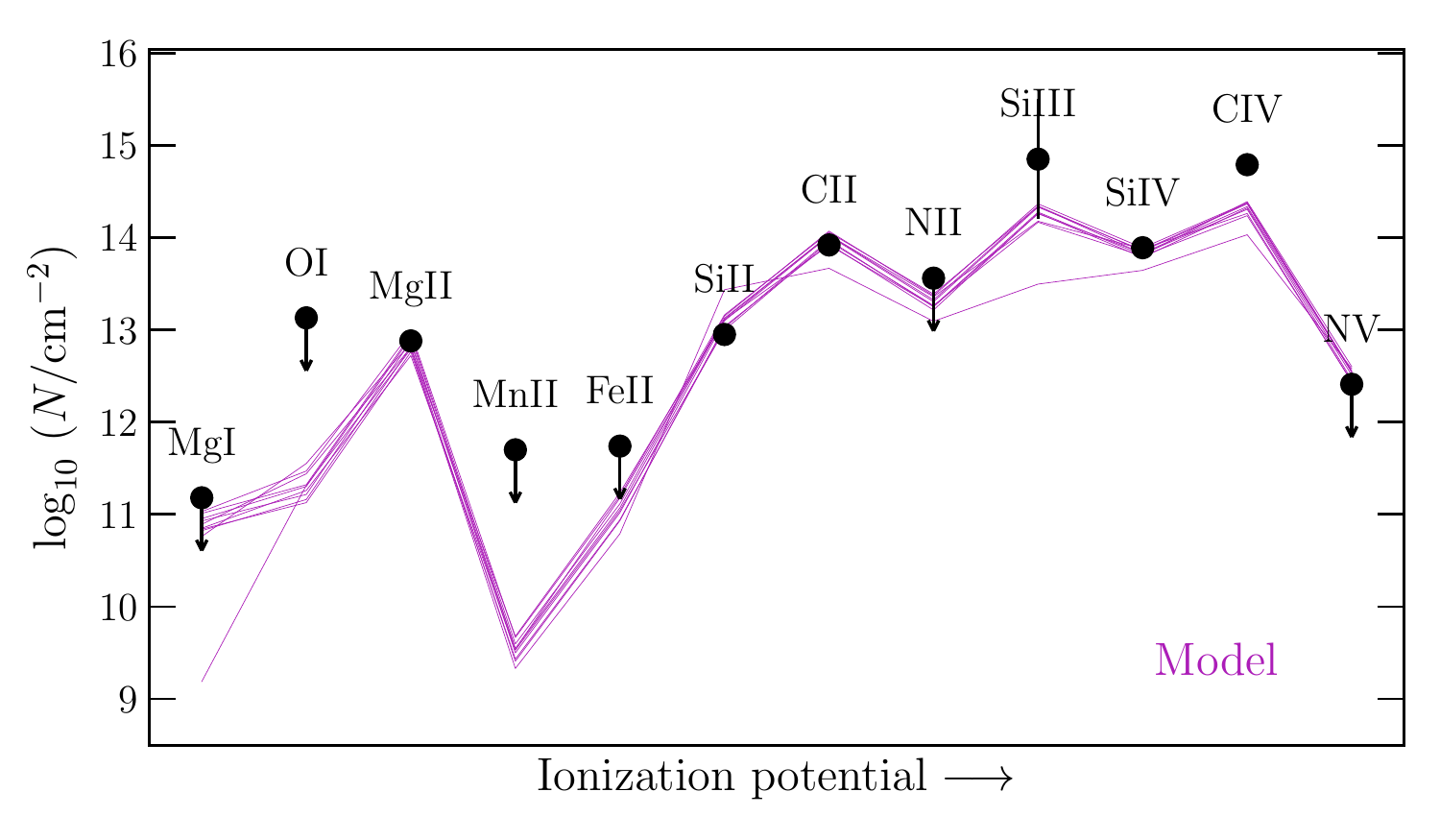}
  \caption{Comparison between the observed column densities (black
    points) and the predicted {\sc Cloudy} column densities for ten
    random MCMC samples (purple) for the All Ions run with HM05.}
  \label{fig:metalmod}
\end{figure}

\begin{deluxetable*}{lccccc}
  \tablecolumns{6}
  \tablecaption{Inferred CGM Metallicities from Photoionization Modeling\label{tab:metallicity}}
  \tablehead{
    \colhead{Run}                &
    \colhead{[Si/H]}             &
    \colhead{$\log U$}           &
    \colhead{$\log n_{\rm H}$}    &
    \colhead{$\log N_{\tiny \HI}$\tablenotemark{a}}  &
    \colhead{$\log N_{\rm H}$}    \\[-2pt]
    \colhead{}                   &
    \colhead{}                   &
    \colhead{}                   &
    \colhead{(cm$^{-3}$)}         &
    \colhead{(\cmsq)}            &
    \colhead{(\cmsq)}
  }
  \startdata
      \cutinhead{HM05 UV Background}\\[-12pt]
      All Ions                  & $-1.5_{-0.3}^{+0.4}$ & $-2.35_{-0.08}^{+0.06}$ & $-2.32_{-0.06}^{+0.08}$ & $17.6_{-0.4}^{+0.3}$    & $20.5_{-0.3}^{+0.3}$ \\
      No {\NV}                  & $-1.8_{-0.1}^{+0.5}$ & $-2.14_{-0.11}^{+0.01}$ & $-2.53_{-0.01}^{+0.11}$ & $17.79_{-0.55}^{+0.04}$ & $20.8_{-0.2}^{+0.2}$ \\
      No {\NV}, {\CIV}          & $-1.7_{-0.2}^{+0.7}$ & $-2.25_{-0.18}^{+0.06}$ & $-2.42_{-0.06}^{+0.18}$ & $17.7_{-0.7}^{+0.1}$    & $20.5_{-0.4}^{+0.4}$ \\
      No {\NV}, {\CIV}, {\SiIV} & $-1.6_{-0.4}^{+0.4}$ & $-2.5_{-0.2}^{+0.2}$   & $-2.2_{-0.2}^{+0.2}$    & $17.6_{-0.5}^{+0.5}$    & $20.5_{-0.4}^{+0.4}$ \\[2pt]
      \cutinhead{HM12 UV Background}\\[-12pt]
      All Ions                  & $-0.43_{-0.75}^{+0.07}$ & $-2.64_{-0.02}^{+0.05}$   & $-2.25_{-0.05}^{+0.02}$   & $16.67_{-0.09}^{+0.65}$ & $19.8_{-0.5}^{+0.5}$ \\
      No {\NV}                  & $-1.66_{-0.02}^{+0.23}$ & $-2.234_{-0.08}^{+0.001}$ & $-2.656_{-0.001}^{+0.08}$ & $17.65_{-0.23}^{+0.04}$ & $20.9_{-0.1}^{+0.1}$ \\
      No {\NV}, {\CIV}          & $-1.80_{-0.07}^{+0.29}$ & $-2.27_{-0.13}^{+0.01}$   & $-2.62_{-0.01}^{+0.13}$   & $17.84_{-0.34}^{+0.08}$ & $21.0_{-0.1}^{+0.1}$ \\
      No {\NV}, {\CIV}, {\SiIV} & $-1.84_{-0.12}^{+0.36}$ & $-2.38_{-0.29}^{+0.08}$   & $-2.51_{-0.08}^{+0.29}$   & $17.92_{-0.42}^{+0.14}$ & $20.8_{-0.3}^{+0.3}$ \\[5pt]
  \enddata

  \tablenotetext{a}{These values are inferred from the MCMC analysis,
    but the value was assumed to be in the range of $15.00 \leq \log
    N_{\tiny \HI} \leq 18.18$ from Table~\ref{tab:columns}.}
\end{deluxetable*}

To avoid unphysically large cloud sizes, we restrict total cloud sizes
to be below 100~kpc and apply this as a flat prior. While this upper
limit allows for small sizes, applying a total cloud size upper limit
of $\sim10$~kpc or below as a prior is too restrictive for the
absorber studied here, resulting in unphysically small total cloud
sizes of $\sim10$~pc or MCMC walkers getting stuck in their initial
positions. This {\it total} cloud size value of $\sim10$~pc is over an
order of magnitude smaller than the {\it component} cloud sizes
estimated by \citet{crighton15}, who found component cloud sizes to
range from $100-500$~pc. This corresponds to a total cloud size of
$600-3000$~pc for a six component absorber (i.e., {\MgII}) or
$900-4500$~pc for a nine component absorber (i.e., {\CIV}). A total
cloud size of 10~pc is also over an order of magnitude smaller than
the total cloud sizes inferred by \citet{lehner16}, who studied a
statistical sample of $z\sim2$ LLSs and pLLSs, where their total cloud
sizes range from $0.07-1862$~kpc (mean of 112~kpc and median of
14~kpc).

Figure~\ref{fig:corner} presents the posterior distributions from the
MCMC modeling for metallicity, [Si/H], ionization parameter, $\log U$,
hydrogen density, $\log n_{\rm H}$, and {\HI} column density, $\log
N_{\tiny \HI}$ for the All Ions run using the HM05 UV background. For
the analysis, 100 walkers were initialized with a burn-in stage of 200
steps and then were run for another 200 steps to determine the final
distributions. Histograms show the distributions for each quantity,
where the vertical lines indicate the maximum likelihood value (thick
lines) with $68\%$ confidence interval uncertainties (thin lines). The
final adopted values for metallicity, ionization parameter, hydrogen
density, {\HI} column density, and total hydrogen column density are
listed in Table~\ref{tab:metallicity}. Our analysis infers a
metallicity of $[{\rm Si}/{\rm H}]=-1.5^{+0.4}_{-0.3}$ when all ions
are included.

A comparison between the measured column densities and ten random MCMC
sample predictions is plotted in Figure~\ref{fig:metalmod} for the All
Ions run with HM05. Black points represent the column densities from
Table~\ref{tab:columns}, where the uncertainties on the points are
mostly smaller than the points themselves. The large uncertainty on
the {\SiIII} line is due to saturation at $v=0$~{\kms}. The purple
lines indicate the ten MCMC samples, which reasonably predict most
column densities. However, the models do not accurately predict
{\CIV}, which is likely due to the fact that the ion traces
intermediate ionization gas. {\CIV} often displays kinematics that are
consistent with both the low ions (e.g., {\MgII} and {\SiII}) and the
high ions (e.g., {\OVI}). While we do not have coverage of {\OVI} for
this absorption system here to further examine this, the intermediate
ionization kinematic behavior is discussed in, e.g., \citet{rudie19}.

Another potential reason for the discrepancy between the measured
{\CIV} column density and the model values is that we include {\NV} in
the analysis, which is a higher ionization transition. In this case,
our assumption of a single phase may break down. To investigate the
dependence of metallicity on the inclusion of the higher ions, we
conduct a test removing {\NV}, {\CIV}, and {\SiIV} from the MCMC
analysis one-by-one. The results of these tests are listed in
Table~\ref{tab:metallicity} for comparison. Overall the resulting
metallicities are all consistent within uncertainties, even between
the ``all ions'' and ``no {\NV}'' runs, which represent the extremes
in metallicity. We examined the comparison between the observed column
densities and predicted {\sc Cloudy} column densities (i.e.,
Figure~\ref{fig:metalmod}) for the ``no {\NV}'' run and found that the
models more accurately predict the {\CIV} column densities, but the
comparison for the rest of the ions does not differ appreciably. For
the ``no {\NV}, {\CIV}'' and ``no {\NV}, {\CIV}, {\SiIV}'' runs, the
random models continue to trace the observed (included) ions similarly
to those plotted in Figure~\ref{fig:metalmod}, but the variation in
the models increases. This latter comparison suggests that {\CIV} and
{\SiIV} provide important constraints on the metallicity estimate,
which is sensible since these ions exhibit kinematic structures that
are similar to the lower ions. The assumption of a single phase may
then be valid in this system. We cannot investigate this further since
it requires partitioning {\HI} between phases, which is difficult with
only the saturated {\Lya} line and we do not wish to over-interpret the
system. The total cloud sizes (${\rm cloud~size}= N_{\rm H}/n_{\rm
  H}$) derived from this modeling and single phase assumption are
reasonable, with 21~kpc for the All Ions run, 69~kpc for ``no {\NV}'',
27~kpc for ``no {\NV}, {\CIV}'', and 16~kpc for ``no {\NV}, {\CIV},
{\SiIV}''. These values are consistent with the $z=2$ LLSs studied by
\citet{lehner16}. Given the results of these tests, we adopt the
metallicity inferred from the ``all ions'' run with HM05 for the
remainder of our analysis.

For comparison to other work that uses the HM12 UV background, we also
modeled the absorption system after changing HM05 to HM12 but keeping
the priors and {\sc Cloudy} grids the same. These results are listed
in Table~\ref{tab:metallicity}. There is a much larger variation in
the metallicity inferred between the four different runs, with the All
Ions run having a higher metallicity than the other runs and all HM05
runs. It appears that the inclusion of {\NV} with HM12 skews the
metallicity toward much higher values because the values decrease once
{\NV} is removed from the analysis. This would not be unexpected since
a multiphase and multicomponent analysis of a $z=0.4$ absorber--galaxy
pair found a higher metallicity for the higher ionization phase
\citep[e.g.,][]{muzahid15}. The downward uncertainties on this HM12
All Ions run are clearly large, indicating that the metallicity is
likely lower, and are consistent with the All Ions run with
HM05. Other than the All Ions run, the HM12 results are all consistent
within uncertainties and consistent with their comparable HM05
runs. The choice of UV background does not significantly affect our
conclusions.

\newpage
\subsection{Metallicity--Azimuthal Angle Relation}
\label{sec:metal-aa}

\begin{figure}[t]
  \centering
  \includegraphics[width=\linewidth]{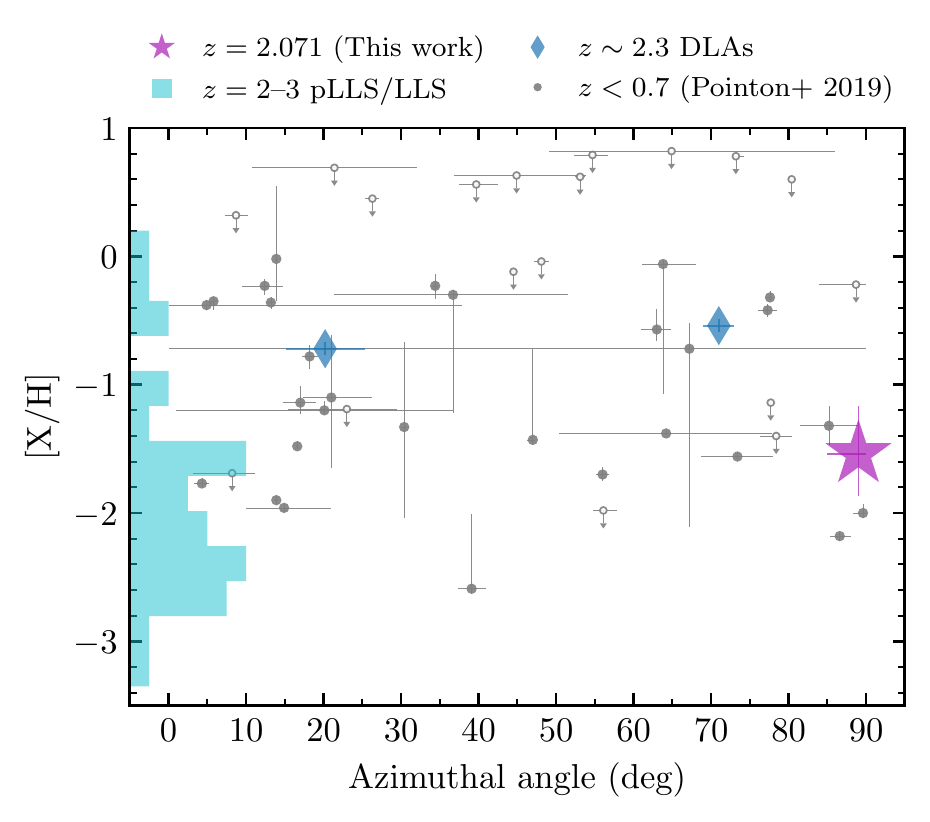}
  \caption{CGM metallicity, $[{\rm X}/{\rm H}]$, as a function of
    azimuthal angle, $\Phi$. Galaxies probed along their projected
    major axis have $\Phi=0^{\circ}$, while galaxies probed along
    their projected minor axis have $\Phi=90^{\circ}$. The
    J1430$+$014, $z=2.071$ minor axis absorber (this work) is plotted
    as the purple star, while the $z\sim2.3$ DLAs from
    \citet{bouche13} and \citet{krogager13} are plotted as the blue
    diamonds. The cyan histogram shows the distribution of pLLS/LLS at
    $z=2-3$ from \citet{crighton13, crighton15} and \citet{lehner14,
      lehner16}. Also plotted for comparison are gray points, which
    represent CGM metallicities for isolated galaxies at $z<0.7$ from
    \citet{pointon19iso}, where solid points indicate metallicity
    measurements and open points indicate upper limits on the CGM
    metallicity.}
  \label{fig:metallicity}
\end{figure}

At low redshift, $z<0.7$, \citet{pointon19iso} found that the total
CGM metallicity along the line-of-sight does not depend on the
orientation at which the quasar probes isolated galaxies within
200~kpc. The authors suggested that the lack of a dependence may
partially be due to studying the CGM during a time at which gas flows
are diminishing in strength. Here we place the J$1430+014$, $z=2.071$
absorber--galaxy pair in the context of both this low redshift survey
and $z=2-3$ metallicity distributions. Figure~\ref{fig:metallicity}
presents the available data sets, which will later be used with our
full program to understand if gas flows are more distinct on this
plane when galaxies are most actively building up their stellar
mass. The \citet{pointon19iso} data are plotted as gray points for
comparison, where upper limits on the metallicity are plotted as open
points. The metallicities for the Pointon et~al.~sample were measured
with the same methods used here, so we can directly compare their
values.

Since the absorber we study here is a pLLS or a LLS (see
Table~\ref{tab:columns} for the adopted $\log N_{\tiny \HI}$ range),
we plot a cyan histogram of the $z=2-3$ pLLS/LLS sample from
\citet{crighton13, crighton15} and \citet{lehner14, lehner16} on the
y-axis for comparison.\footnote{Also see \citet{fumagalli16}. Note
  that the authors of that study use the HM12 ionization
  background. At low redshifts, HM12 results in metallicities that are
  roughly 0.3~dex higher than the HM05 background we use
  here. Table~\ref{tab:metallicity} lists HM12 results for comparison
  to \citet{fumagalli16} but we refrain from doing the comparison here
  as our focus is on the large body of work at low redshift, which
  mostly uses HM05.} In most cases, host galaxies have not yet been
identified for that sample and where they have been identified, there
is no published morphology information. Also plotted are two
$z\sim2.3$ DLAs (blue diamonds) with measured azimuthal angle
information \citep{bouche13, krogager13}. The J1430$+$014, $z=2.071$
absorber studied here is plotted as the purple star. The figure is
currently sparsely populated by $z=2-3$ absorber--galaxy pairs, but we
aim to expand the sample with our KCWI program. Recall that the KBSS
\citep{rudie12} also examines the CGM over this redshift range, but
does not currently provide the CGM metallicities nor azimuthal angles.

The J1430$+$014, $z=2.071$ absorber is consistent with the range of
metallicities found at low redshift, $z<0.7$, although it is on the
lower end of the range. This is expected since the metallicity of
pLLSs/LLSs evolves from lower metallicities at $z=2-3$ to higher
metallicities at $z<1$ \citep{lehner16}. Compared to the $z=2-3$
pLLSs/LLSs, the J1430$+$014 absorber is slightly more metal-rich than
the peak in the distribution at $[{\rm X/H}]\sim-2$. Thus it is likely
that the gas was previously in the galaxy at some point, where an
outflow is a plausible source of the observed gas in this
sightline. We expand on this further in Section~\ref{sec:outflowmod},
where we model the gas as an outflowing cone.

Thus far there are only three $z=2-3$ sightlines where both CGM
metallicities and galaxy morphologies have been measured. Compared to
the minor axis DLA at $D=6$~kpc from \citet{krogager13} and the major
axis DLA at $D=26$~kpc from \citet{bouche13}, our absorber--galaxy
pair is more metal-poor. However, DLAs are known to be more metal-rich
than pLLSs/LLSs since the background quasar is thought to probe the
disks or near the disks of the host galaxies, so this result is not
unexpected \citep[e.g.,][]{lehner16}. From these three points, it is
unclear if the separation between metal-poor, major axis gas and
metal-rich, minor axis gas will be clear at $z=2-3$. \citet{peroux20}
studied the CGM metallicity--azimuthal angle plane in the EAGLE and
IllustrisTNG simulations and suggested that there should not be a clear
trend at this redshift due to galaxies not having enough time to
adequately pollute their CGM with metal-enriched gas. In order to best
examine this, we need a large sample of absorber--galaxy pairs with
similar {\HI} column density ranges (i.e., pLLSs/LLSs). We plan to
further explore the CGM metallicity--galaxy relation with a larger
sample of $z=2-3$ absorber--galaxy pairs in the future.

\subsection{Outflow Modeling}
\label{sec:outflowmod}

Assuming the host galaxy has an edge-on morphology, the kinematics and
metallicity of the observed gas and its location relative to the host
galaxy suggest the gas is plausibly outflowing material. If we then
assume outflowing material, we can use the combination of {\MgII}
kinematics and galaxy geometry in the conical model detailed by
\citet{gauthier12} to constrain the outflow velocity and opening
angle.\footnote{While the \citet{gauthier12} model can explain the
  observed velocity space in the absorption profiles, it does result
  in velocity gradients that are too large compared to momentum-driven
  outflow models (e.g., supernovae feedback). However, this simple
  model results in outflow characteristics that are comparable to the
  simple outflow models of \citet{bouche12}, \citet{martin19}, and
  \citet{schroetter16, schroetter19}, so we use the model here for
  comparison to other work.} Figure~1 in their paper best illustrates
the model, which we briefly describe here. The outflow is defined as
an outwardly expanding polar cone emanating from the galaxy center
with a full angular span of $2\theta_0$. The quasar sightline enters
the outflowing cone at a height above the galaxy disk $z_1$ and exits
at $z_2$, both of which depend on the galaxy geometry and outflow
opening angle, where we require that $z_1 < z_2$. The galaxy has an
inclination, $i$, and azimuthal angle, $\Phi$, which is the angle
between the projected galaxy disk and the background quasar
sightline. The quasar is located at an impact parameter, $D$, from the
galaxy. All galaxy properties are defined in Table~\ref{tab:galaxy}
for the absorber--galaxy pair studied here.

The relation between the half opening angle, $\theta_0$, and
$z_{[1,2]}$, which is the outflow disk (i.e., a slice through the
outflow cone parallel to the galaxy) at heights $z_1$ and $z_2$, is
\begin{equation} \label{eq:zheight}
  z_{[1,2]} \tan \theta_{0}=D \sqrt{1+\sin ^{2} \phi_{[1,2]} \tan ^{2}
    i}\left(\frac{\cos \Phi}{\cos \phi_{[1,2]}}\right).
\end{equation}
The position angle between the quasar sightline and the outflow
disks projected on the sky, $\phi_{[1,2]}$, is defined by
\begin{equation} \label{eq:phi}
  \tan \phi_{[1,2]}=\frac{D \sin \alpha-z_{[1,2]} \sin i}{D \cos \alpha}.
\end{equation}
Using Equations~\ref{eq:zheight} and \ref{eq:phi} we can then
calculate the angle, $\theta$, along the quasar sightline for a height
$z$, where $z_1 \leq z \leq z_2$. Finally, the outflow speed, $V_{\rm
  out}$, of a gas cloud moving outward from the galaxy at height $z$
is related to the line-of-sight velocity, $v_{\rm los}$, and $\theta$
with
\begin{equation} \label{eq:vout}
  V_{\rm out}=\frac{v_{\mathrm{los}}}{\cos j}, \text { where } j=\sin
  ^{-1}\left(\frac{D}{z} \cos \theta\right),
\end{equation}
and $\theta \leq \theta_0$. $V_{\rm out}$ is the radial outflow speed
from the galaxy, which varies with $z$-height above the galaxy, and
depends on the line-of-sight velocity and the half opening angle of
the outflow cone for a given $z$-height. This value is not necessarily
constant.

From all of this, the input values for the model include absorption
velocity bounds $v_{\rm los,1}$ and $v_{\rm los,2}$, galaxy properties
$D$, $i$, and $\Phi$, and the range of possible $z$-heights from 0~kpc
to 1000~kpc. The model assumes that the {\MgII}-absorbing gas fills
the outflow cone such that the velocity bounds on absorption represent
the physical bounds on the cone. These bounds are $v_{\rm
  los,1}=56$~{\kms} and $v_{\rm los,2}=-151$~{\kms} and were
determined by identifying the velocities relative to $z_{\rm gal}$ at
which the absorption profile recovers to within $1\%$ of the continuum
(see Table~\ref{tab:ews} and Section~\ref{sec:absorption}). The
line-of-sight velocities are assumed to increase smoothly as the
quasar sightline passes through the outflow cone towards larger
$z$-heights, which implies that the outflow cone is pointed towards
the observer rather than away.

\begin{figure}[t]
  \includegraphics[width=\linewidth]{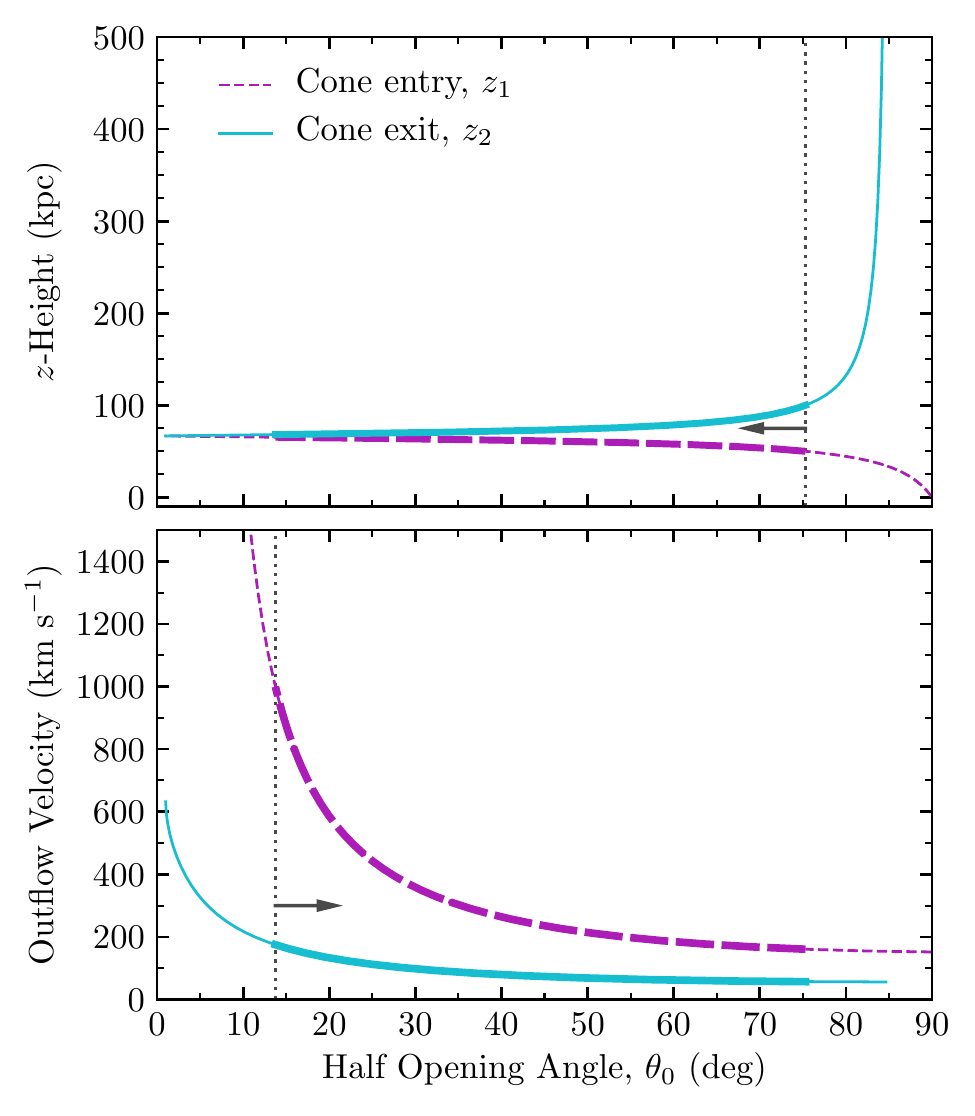}
  \caption{Outputs from modeling the {\MgII} absorption as a polar
    outflowing cone: $z$-heights, half opening angles, and outflow
    velocities. The quasar sightline enters the outflowing cone at
    $z_1$ (dashed purple) and exits the cone at $z_2$ (cyan), where
    $z_1 < z_2$. Vertical dashed lines and thicker curves indicate
    allowed half opening angles, beyond which the $z$-heights or
    outflow velocities asymptote to unrealistically large values. The
    gas is decelerating for all half opening angles.}
  \label{fig:outflowmod}
\end{figure}

Figure~\ref{fig:outflowmod} presents the allowed $z$-heights, half
opening angles, and outflow velocities for the J1430$+$014, $z=2.071$
absorber--galaxy pair, assuming the host galaxy is a single object
with an edge-on inclination. The purple dashed line represents the
location at which the quasar sightline enters the outflowing cone,
while the solid cyan line is the exit location. The model requires
that $z_1 < z_2$, which is true in the upper panel for all half
opening angles. From the lower panel, the outflow velocity decreases
from $z_1$ to $z_2$ for all half opening angles, indicating that the
gas is decelerating as it outflows. With these curves, we can then
constrain the allowed half opening angle, $\theta_0$, and therefore
the outflow velocity, $V_{\rm out}$. Considering only the
absorber--galaxy geometry, the minimum half opening angle allowed
depends on the azimuthal angle between the projected galaxy disk and
the quasar sightline, $\Phi=89^{\circ}$. In order for the sightline to
pass through {\MgII}-absorbing gas, the half opening angle must be at
least the difference $\theta_0 = 90^{\circ}-\Phi = 1^{\circ}$ (recall
that $\Phi=90^{\circ}$ is defined as the projected galaxy minor axis,
which defines $\theta_0=0^{\circ}$ for an edge-on galaxy). Because the
{\MgII} absorption spans the galaxy systemic velocity, i.e., the gas
is both redshifted and blueshifted, we cannot constrain the maximum
opening angle allowed based on geometry alone.

Further considering the distributions of the model parameters in
Figure~\ref{fig:outflowmod}, we find that the $z$-heights asymptote
beyond $\theta_0\sim80^{\circ}$. At these large $\theta_0$ values, the
quasar sightline does not exit the cone beyond
$\theta_0\simeq85^{\circ}$ because the opening angle is large enough
for the edge of the cone to be parallel to the quasar sightline. Given
this, we imposed a maximum allowed $z$-height of 100~kpc, which was
chosen to roughly correspond to the virial radius of a $\log (M_{\rm
  h}/M_{\odot})=12.0$ galaxy at this redshift. The vertical dashed
line in the upper panel indicates the half opening angle corresponding
to this limit, $\theta_0=75^{\circ}$. Similarly, the lower panel shows
that the outflow velocities asymptote to large values below
$\theta_0\sim10^{\circ}$, which is unrealistic since outflows are
generally found to be collimated and have velocities on the order of
several hundred {\kms} \citep[e.g.,][]{bordoloi11, bordoloi14,
  bouche12, kacprzak14, lan14, rubin-winds14, schroetter16,
  schroetter19}.\footnote{However, down-the-barrel observations by
  \citet{law12} suggest that $z=2$ outflows may not yet be collimated
  because the host galaxy disks are still unstable to rotation
  \citep[also see][]{steidel10}, and this may also be true in
  simulations \citep[e.g.,][]{nelson19}. A larger sample is required
  to better investigate this.} Using stacked galaxy spectra,
\citet{steidel10} suggested that outflows at $z=2$ have velocities on
the order of $V_{\rm out}\simeq800$~{\kms}. We therefore constrain the
lower opening angle value using a maximum outflow velocity of
1000~{\kms} and this is indicated by the vertical dotted line in the
bottom panel of Figure~\ref{fig:outflowmod}.

The allowed values from the model are then $14^{\circ} \leq \theta_0
\leq 75^{\circ}$ and are further indicated as the thick curves in
Figure~\ref{fig:outflowmod}. The average outflow velocities
corresponding to these opening angles (i.e., the average between the
purple and cyan curves) are $V_{\rm out}\sim588$~{\kms} for
$\theta_0=14^{\circ}$ down to $V_{\rm out}\sim109$~{\kms} for
$\theta_0=75^{\circ}$. The velocities could be as large as 1000~{\kms}
or as small as 57~{\kms}. Unfortunately these cannot be constrained
further with the geometry of the system. If we account for the
uncertainty in the galaxy redshift that results from the apparent
redshift of the {\Lya} emission, $\pm78$~{\kms}, we find slight
changes to the model results. When a larger {\Lya} correction is
applied, the absorption still spans the galaxy systemic redshift, the
outflow half opening angles could be as small as $9^{\circ}$, and the
gas may accelerate slightly as it moves further from the galaxy, but
the velocities do not change significantly. If a smaller {\Lya}
correction is applied, the absorption then shifts to being completely
blueshifted from the galaxy systemic redshift by -105~{\kms} and the
geometry of the system constrains the half opening angles to
$1^{\circ}\leq\theta_0\leq5^{\circ}$. These opening angles are highly
collimated in contrast to the simulations and other observations at
this epoch and may be unlikely for a star-forming galaxy with no
obvious AGN activity to the limits of the data.

\begin{deluxetable}{crrcc}
  \tablecaption{Outflow Modeling Results\label{tab:Mout}}
  \tablecolumns{5}
  \tablewidth{\linewidth}
  \tablehead{
    \colhead{$\theta_0$\tablenotemark{a}} &
    \colhead{$V_{\rm out}$} &
    \colhead{$t_{\rm out}$\tablenotemark{b}} &
    \colhead{$\dot{M}_{\rm{out}}$~\tablenotemark{c}} &
    \colhead{$\eta$~\tablenotemark{c}} \\[-2pt]
    \colhead{(deg)}                 &
    \colhead{(\kms)}                &
    \colhead{(Myr) }                &
    \colhead{(M$_{\odot}$~yr$^{-1}$)} &
    \colhead{}                      
  }
  \startdata
  \cutinhead{Maximum $V_{\rm out}$}\\[-12pt]
  14 & 1000 &   65 & 88 &  2.3 \\
  75 &  161 &  404 & 77 &  2.0 \\
  \cutinhead{Average $V_{\rm out}$\tablenotemark{d}}\\[-12pt]
  14 &  588 &  111 & 51 &  1.4 \\
  75 &  109 &  597 & 52 &  1.4 \\
  \cutinhead{Minimum $V_{\rm out}$}\\[-12pt]
  14 &  175 &  371 & 15 &  0.4 \\
  83 &   57 & 1140 & 27 &  0.7 \\[2pt]
  \enddata

  \tablenotetext{a}{The lower bound is constrained with a maximum
    allowed $V_{\rm out}$ while the upper bound is constrained with a
    maximum allowed $z$-height (i.e., impact parameter).}

  \tablenotetext{b}{Timescale to $D=66$~kpc given a constant $V_{\rm
      out}$.}

  \tablenotetext{c}{The values reported in these column are considered
    upper limits due to the uncertainty in the outflow and metallicity
    modeling.}

  \tablenotetext{d}{Values calculated using the mean $V_{\rm out}$
    between the $z_{[1,2]}$ curves at the given $\theta_0$.}

\end{deluxetable}

\subsection{Outflow Timescales, Mass Outflow Rates,\\ \& Mass Loading Factors}

Having assumed an edge-on galaxy morphology and modeled the outflow
velocities and half opening angles, we can characterize the impact of
the outflows on the CGM. Although the models suggest the gas is
decelerating, we assume a constant outflow velocity and estimate the
time it takes for the observed gas to reach $D=66$~kpc from the
galaxy. The values are tabulated in Table~\ref{tab:Mout} for various
outflow constraints. At the lower bound for the half opening angle,
the outflow velocity is estimated to be $V_{\rm
  out}=588\pm413$~{\kms}, which results in an outflow timescale of
$t_{\rm out}\sim111_{-46}^{+260}$~Myr or an ejection redshift of
$z_{\rm eject}\sim2.15$. The maximum outflow timescale comes from the
upper bound on the half opening angle, where $V_{\rm
  out}=109\pm52$~{\kms}, $t_{\rm out}\sim597_{-193}^{543+}$~Myr, and
$z_{\rm eject}\sim2.55$. These timescales decrease if the gas is
instead assumed to be decelerating as it moves outwards from the
galaxy.

Despite the large range of allowed half opening angles and outflow
velocities of the potential outflowing gas, we can also estimate mass
outflow rates and mass loading factors to understand how much material
the galaxy is being ejected into the CGM. From \citet{bouche12} and
\citet{schroetter16, schroetter19}, the mass outflow rate is roughly
\begin{equation}
  \frac{\dot{M}_{\mathrm{out}}}{0.5~\mathrm{M}_{\odot}~\mathrm{yr}^{-1}} \approx
  \frac{\mu}{1.5}~
  \frac{N_{\rm H}}{10^{19}~\mathrm{cm}^{-2}}~
  \frac{D}{25~\mathrm{kpc}}~
  \frac{V_{\mathrm{out}}}{200~\mathrm{km}~\mathrm{s}^{-1}}~
  \frac{\theta_{0}}{30^{\circ}},
\end{equation}
where $\mu$ is the mean atomic weight of 1.4. We estimated the total
hydrogen column density, $N_{\rm H}$, from the photoionization
modeling in Section~\ref{sec:metallicity} for the ``all ions'' run,
with $\log N_{\rm H}=20.5$~{\cmsq}. Then the mass loading factor is
\begin{equation}
  \eta=\frac{\dot{M}_{\mathrm{out}}}{\mathrm{SFR}}.
\end{equation}
Table~\ref{tab:Mout} details the mass outflow rates and mass loading
factors for the various opening angle and outflow velocity
constraints. The mass outflow rate is roughly $\dot{M}_{\rm
  out}\sim52\pm37$~M$_{\odot}$~yr$^{-1}$, corresponding to mass
loading factors of $\eta\sim1.4\pm1.0$. These values are likely upper
limits given the galaxy's assumed edge-on inclination, the uncertainty
in the total hydrogen column density from the metallicity modeling,
the high likelihood that the outflow has swept up additional CGM
material along the way, and that the values are the maximum values
obtained if a larger or smaller {\Lya} redshift correction defined by
the uncertainty on the redshift is applied.

These mass outflow rates are roughly an order of magnitude larger than
most values reported by the MEGAFLOW survey at $z\sim1$
\citep{schroetter16, schroetter19}. While this may be due to more
energetic outflows at $z\sim2$ (see discussion in the next section),
the measurement uncertainties are also important. In addition to the
degeneracies in modeling the outflow half opening angles and
velocities, there is a large uncertainty involved in measuring the
total hydrogen column density. MEGAFLOW convert their {\MgII}
equivalent widths to $N_{\tiny \HI}$ using the \citet{menard09}
relation and assume that the ionized gas contribution is negligible,
such that $N_{\tiny \HI}\approx N_{\rm H}$. Our $N_{\rm H}$ is
inferred from the photoionization modeling and accounts for the
possibility that hydrogen is also associated with gas that is higher
ionization than {\MgII}, such as {\CIV} which is measured for our
absorber. Compared to MEGAFLOW, our {\MgII} absorber has a smaller
equivalent width, but our total hydrogen column density is larger than
most objects in their sample. This suggests that mass flow rates
determined from statistical relations with large intrinsic scatter
should be viewed with caution.

\section{Discussion}
\label{sec:discussion}

As detailed in the previous sections, the combination of absorbing gas
observed along the minor axis of an assumed edge-on galaxy with broad
kinematics symmetrically distributed around $z_{\rm gal}$ suggests an
outflow origin despite the large impact parameter, while the
relatively metal-enriched metallicity suggests the outflow likely
entrained lower metallicity gas on its way out from the galaxy. This
is not unexpected since even at low redshift, $z<1$, outflows appear
to influence the CGM out to at least 50~kpc \citep{bordoloi11} and
potentially further \citep[100~kpc;][]{lan18, schroetter19}. Outflows
at $z=2$ are more energetic with higher velocities and mass loading
factors \citep[e.g.,][]{sugahara17}. It could be expected that the
material ejected from these outflows could reach larger galactocentric
distances than measured at low redshifts, although the outflows would
be competing with increased accretion rates as well. There is already
potential evidence of outflowing material reaching out to 180~kpc at
$z=2$ \citep{turner14, turner15, turner17}. These large impact
parameters also result in long timescales and sufficiently early
ejection redshifts for the SFR measured to be a lower limit.

\subsection{Galaxy Morphology}
\label{sec:galmorph}

Our determination that the observed CGM gas has an outflow origin
relies partially on the galaxy morphology, which can be difficult to
quantify at $z=2$ due to the clumpy nature of galaxies at this epoch
\citep[e.g.,][and references therein]{guo15}. In a shallow {\it
  HST}/ACS image covering rest-frame UV wavelengths (F625W), we find
two bright clumps with fainter material, but still a considerable
amount of continuum, distributed in a rough plane perpendicular to the
quasar sightline. These clumps likely correspond to separate star
formation regions in the object given the wavelength coverage. This
morphology is similar to both the ``double'' and ``chain'' galaxy
morphologies studied in the {\it HST} Ultra Deep
Field \citep[e.g.,][]{elmegreen07}. Galaxies with two clumps are
common at this epoch \citep[e.g.,][]{shibuya16} but the physical
origin of these clumps is still debated in the
literature. \citet{shibuya16} suggest that objects with this two-clump
morphology are only formed via a violent disk instability whereas
\citet{ribeiro17} suggests that they are formed via major
mergers. \citet{fisher17} studied DYNAMO galaxies identified as low
redshift analogs of star-forming galaxies at $z=1-3$. The authors
found that multiple clumps that are clustered in the disks of these
galaxies appear as fewer, larger clumps in the lower resolution
imaging that is common at higher redshift. This further complicates
the interpretation of double clump morphologies, especially in shallow
imaging.

As these works show, it is difficult to determine if two-clump
morphologies are mergers or disk galaxies from morphological or
kinematic classification alone \citep{shapiro08, hung15}. To best
determine whether our object is a merger or not, we would require both
morphology and kinematics \citep{rodrigues17}. We do not currently
have kinematics for this object to better classify its morphology
since the spaxels on the medium slicer for KCWI have sizes on the
order of the size of $z\sim2$ galaxies. Regardless, \citet{simons19}
discussed that even with both photometry and kinematics, it may still
be difficult to distinguish between a rotating disk and a merger at
$z\sim2$. This is because merging galaxies can have kinematic
properties that match several criteria used to identify disks at high
redshift \citep[e.g., the five disk criteria comparing kinematic axes,
  velocity dispersions, and velocity gradients used
  by][]{wisnioski15}.

The separation of the two clumps in our object may be a distinguishing
property to determine if the object is best modeled as a major merging
system or a single clumpy galaxy. The two clumps are separated by only
$0\farcs23$, corresponding to 1.9~kpc at this redshift. This
separation is smaller than the typical half light radius of
star-forming galaxies at $z\sim2$ \citep[$\sim2.5-3$~kpc;][]{allen17}
and may indicate that the two clumps are part of a single
object. \citet{vanderwel14} studied the sizes of both star-forming and
quiescent galaxies and found similar sizes as \citet{allen17} for
star-forming galaxies, but smaller sizes for quiescent galaxies on the
order of $\sim1$~kpc at $z\sim2$. It is possible that the clumps in
our object are two spheroids in the process of undergoing a major
merger, but the object is star-forming and the spatial resolution of
KCWI does not allow us to determine separate SFRs of the clumps. If
this object is a merger, then we might reasonably assume that the star
formation rate should be elevated relative to the star formation main
sequence at this redshift. The unobscured SFR for this galaxy,
${\rm SFR}_{\rm FUV}=37.8$~M$_{\odot}$~yr$^{-1}$, is roughly on the main
sequence for $z\sim2$ galaxies with stellar masses of
$\log(M_{\ast}/M_{\odot})\sim10$ \citep[e.g.,][]{daddi07, speagle14},
so a merger seems less likely. While \citet{pearson19} suggest that a
merger does not necessarily result in highly elevated star formation
rates, they do find that the merger fraction increases the further a
galaxy is away from the star formation main sequence. The main
sequence SFR of our object indicates that a major merger morphology
may have a lower probability than a single clumpy galaxy morphology.

Fortunately, simulations can provide some insight into this problem
where observations currently struggle. \citet{simons19} studied
theoretical merger fractions and the probability that mergers are
classified as disks with the \citet{wisnioski15} disk criteria. The
authors estimated that 5-15\% of disks identified at these redshifts
with kinematic criteria are actually mergers. More relevant to the
work presented here is that \citet{simons19} found that the fraction
of $z\sim2$ Illustris galaxies with a companion within a 2D projected
separation of $\sim2$~kpc (i.e., the separation of the two clumps in
our object) is at most $2-3\%$ for all merging systems, $\sim1\%$ for
low mass major mergers, and nearly $0\%$ for high mass major
mergers. Given the low probability of merging galaxies with this
separation, the separation of the two clumps being smaller than
typical star-forming galaxies, and the low fraction of merging
galaxies on the star formation main sequence (where this object lies
at this redshift), we make the assumption that the object we observed
is a single, clumpy, edge-on galaxy.

\subsection{Outflow Kinematics}
\label{sec:okin}

Assuming the galaxy is a clumpy disk with an edge-on inclination from
the discussion in the previous section, and having determined that the
galaxy is star-forming with ${\rm SFR}_{\rm
  FUV}=37.8$~M$_{\odot}$~yr$^{-1}$, we modeled the observed gas as an
outflow. The kinematics of the metal lines, where absorption is
detected, all have similar velocity distributions and are all
reminiscent of the profiles found in the Milky Way Fermi bubbles. That
is, the gas is roughly centered on the galaxy systemic redshift,
$z_{\rm gal}$, for this edge-on galaxy and this is where the bulk of
the absorption lies. There are additional roughly symmetric higher
velocity components out to $v\pm50-150$~{\kms} for the lower
ionization lines. Furthermore, the full velocity width of {\MgII},
$\Delta v\sim210$~{\kms}, is consistent with the large velocity
separation tail in the pixel-velocity two-point correlation function
of the bluer (more highly star-forming) galaxies probed along their
projected minor axes in \citet{magiicat5}. The authors concluded these
long tails were consistent with biconical outflows with large velocity
dispersions. All of these corroborating inferences imply that the
observed gas fills an outflowing cone, where absorption velocity
bounds define the physical bounds on the cone. This commonly invoked
scenario results in reasonable average outflow velocities ($109 <
V_{\rm out} < 588$~{\kms}) at half opening angles ($14^{\circ} <
\theta_0 < 75^{\circ}$) that are comparable with the lower redshift
quasar absorption line technique \citep[e.g.,][]{gauthier12,
  bordoloi14-model, kacprzak14, schroetter16, schroetter19, martin19},
down-the-barrel observations \citep[e.g.,][]{weiner09, coil11,
  martin12, rubin-winds, rubin-winds14}, and azimuthal angle
distributions \citep[e.g.,][]{bouche12, kcn12, kacprzak15, lan14}.

These outflow characteristics are also consistent with the recent
IllustrisTNG TNG50 simulations, where \citet{nelson19} examined
supernovae- and black hole-driven outflows at $z>1$. They found that
$T\sim10^{4.5}$~K gas (i.e., {\MgII}) is confined to maximum outflow
velocities of $V_{\rm out}<500$~{\kms}, consistent with values we
measure here. Collimation of the outflows was observed by $z=1$, where
the gas was located along the galaxy minor axis with half opening
angles of $\theta_o=40^{\circ}-50^{\circ}$, although the authors
suggested collimation may not be strongly present at $z=2$ \citep[also
  see][]{steidel10, law12}. The maximum outflow velocity we
constrained the model to, 1000~{\kms}, corresponds to very small half
opening angles. This might indicate that the highest velocity gas
possible for this system should be collimated in contrast to previous
findings that found high velocities with low collimation
\citep[e.g.,][]{steidel10}. It is difficult to further explore this
scenario with the single absorber--galaxy pair studied here,
especially since the modeled galaxy morphology parameters provide
little constraint on the opening angles, i.e., the galaxy is assumed
to be nearly perfectly edge-on ($i=85^{\circ}$) and probed along the
minor axis ($\Phi=89^{\circ}$). A wider variety of galaxy morphologies
will better constrain the typical $z=2$ outflow
properties. Furthermore, a low collimation of outflows at these
redshifts would mean that galaxy morphology is a less important gas
flow diagnostic than at low redshift. In the case studied here, the
uncertainty in the host galaxy's morphology may not be important. This
can be better tested with our forthcoming sample.

The timescales and ejection redshifts for the outflowing gas to reach
the observed impact parameter of $D=66$~kpc, $t_{\rm out}=109-588$~Myr
and $z_{\rm eject}=2.15-2.55$, are consistent with the timescales
required for outflowing gas to recycle back onto galaxies (e.g.,
\citealp{oppenheimer10, angles17}; although see
\citealp{keller19}). They are also about $1-2$ orders of magnitude
larger than the lifetime of a starburst
\citep[e.g.,][]{thornley00}. Given that the gas we observed is cool,
photoionized material (at least for ions such as {\MgII}, with
$T\sim10^{4.5}$~K), it may be surprising that the gas has not yet been
destroyed. Furthermore, \citet{crighton15} found that these clouds
tend to be small, on the order of a few hundred parsecs \citep[and
  even as small as $\sim10$~pc;][]{weakII}, sometimes characterized as
circumgalactic ``mist'' or ``fog'' in simulations
\citep[e.g.,][]{liang20, mccourt18}, and so may be susceptible to
being destroyed in an outflowing environment. Either the gas is being
reformed as cool material condensing out of a hot halo
\citep[e.g.,][]{simcoe06, fraternali13}, or the clouds are embedded in
progressively hotter phases \citep[e.g.,][]{stern16}. We support this
latter scenario since we find multiphase gas with similar, but
increasingly broader kinematics as the ionization state
increases. This is particularly noticeable with the Si transitions:
{\SiII}, {\SiIII}, and {\SiIV}.

The final parameters obtained from the outflow modeling, mass outflow
rates and mass loading factors, also present an interesting picture of
the outflowing gas from this galaxy. These values are $\dot{M}_{\rm
  out}<52\pm37$~M$_{\odot}$~yr$^{-1}$ and $\eta<1.4\pm1.0$, implying
an energetic and efficient outflow. In the {\sc Simba} simulations
\citep{simba}, a typical star-forming galaxy similar to the galaxy we
present here has a mass outflow rate of $\dot{M}_{\rm
  out}\sim70$~M$_{\odot}$~yr$^{-1}$ and $\eta\sim0.8$, given a ${\rm
  SFR}\sim85$~M$_{\odot}$~yr$^{-1}$. Our mass outflow rate and mass
loading factor are comparable to the simulated galaxies as well as
mass loading factors measured via a variety of estimators at low
redshift \citep[e.g.,][]{weiner09, nicha19}.

Compared to most measurements from the $z\sim1$ quasar absorption line
sample MEGAFLOW \citep{schroetter16, schroetter19}, who study stronger
{\MgII} absorbers with $W_r(2796)\geq0.5$~{\AA}, our mass outflow
rates are roughly an order of magnitude larger. This is likely due to
the combination of less energetic/efficient outflows on average
towards lower redshifts and the large uncertainties involved in
measuring the total hydrogen column density. The few systems in their
sample with comparable outflow rates to ours are also mostly located
at $D>50$~kpc. The values we measure are also larger than the
\citet{crighton15} $z\sim2.5$ outflow ($\dot{M}_{\rm
  out}=5$~M$_{\odot}$~yr$^{-1}$) at $D=50$~kpc, although the authors
do not have galaxy morphologies in order to more accurately model the
gas flows. A more comparable absorber--galaxy pair is the metal-rich
outflow at $z\sim0.4$, $D=163$~kpc from \citet{muzahid15}, which has
an ultra-strong {\OVI} absorption profile with $\dot{M}_{\rm
  out}\sim54$~M$_{\odot}$~yr$^{-1}$. Most of these results are for
absorbers located at fairly large impact parameter, which likely
introduces additional uncertainty into the measurement since the gas
could be destroyed as it moves out from the galaxy and the outflow
velocities are likely not constant. Furthermore, the outflow likely
sweeps up additional material as it plows through the CGM, so our
values are better characterized as upper limits. Regardless, obtaining
more $z=2-3$ measurements will better constrain the outflow properties
to compare to the large number of $z\lesssim1$ systems.

There is still a non-zero probability that the galaxy is actually a
major merging system. In this case, we might still observe outflowing
material along the quasar sightline even if the J1430$+$014, $z_{\rm
  gal}=2.0711$ object is a major merger. \citet{rupkemakani} studied a
galactic wind in emission from a $z\sim0.459$ galaxy merger that is
highly star-forming (${\rm SFR}~\sim100-200$~M$_{\odot}$~yr~$^{-1}$;
much further from the star formation main sequence than our
galaxy). They identify bipolar outflow bubbles that are oriented
perpendicular to the projected major axis of the object's stellar
component and most prominent tidal tail. Even though the projected
major axis of a merging system would be expected to change over time,
the authors still found evidence for two outflow episodes as long as
0.4~Gyr and 7~Gyr ago with velocities $>100$~{\kms} within this
emission structure. These timescales and velocities are consistent
with the values measured for the J1430$+$014 object assuming a single
edge-on galaxy.

\subsection{CGM Metallicity}
\label{sec:metaldisc}

The inferred total CGM metallicity along the line-of-sight for the
J1430$+$014 absorber, ${\rm [Si/H]}=-1.5$, is in the more metal-rich
half of the pLLS/LLS distribution at $z=2-3$ \citep{lehner16}, which
may favor an outflow scenario. With particle tracking, \citet{hafen19}
examined the metallicities of accreting and outflowing gas at $z=2$ in
the FIRE simulations and found that the median metallicity of
outflowing gas is typically $\sim1$~dex larger than IGM accretion,
with $\log(Z/Z_{\odot})\gtrsim-1.0$ for $\log (M_{\rm
  h}/M_{\odot})>11$. The mass--metallicity relation for $z\sim2$
galaxies indicates that $>L_{\ast}$ galaxies have gas-phase
metallicities greater than $-0.7$~dex below solar \citep{steidel14,
  kacprzak15mmr, sanders15}, and so any material ejected recently from
the galaxy might be expected to have a similar metallicity. These
values are more metal-rich than the metallicity we inferred for our
absorber but we do not know the galaxy metallicity to determine how
different the two values are.

The distribution of pLLS/LLS at $z\sim2$ suggests that CGM
metallicities above ${\rm [X/H]}>-1.0$ are rare for systems like ours
\citep[$\sim13\%$;][]{lehner16}. However, outflows are ubiquitous
around star-forming galaxies at $z=2-3$ \citep[e.g.,][]{steidel10} and
these outflows must be probed by quasar sightlines, so the lower CGM
metallicity compared to the typical ISM metallicity is puzzling. Using
the EAGLE and IllustrisTNG TNG50 simulations, \citet{peroux20} studied
CGM metallicities as a function of azimuthal angle and found a
constant metallicity of $\sim-1.8$ for $z=2.4$ galaxies, which is
slightly lower than our inferred value. Even though outflows are
present in their simulations, there is no increase in the metal
content for gas along the minor axis. The authors suggested that this
was because galaxies at this epoch have not yet had enough time to
enrich their CGM. Furthermore, low redshift observational studies have
not yet demonstrated that minor axis gas, which should be dominated by
outflows, has a CGM metallicity similar to the host galaxy ISM
metallicity. At $z\sim0.3$ \citet{kacprzak19b} found that total CGM
metallicities for minor axis sightlines are lower than ISM
metallicities, with values on average $-1.23$ dex below the
ISM. \citet{peroux16} found a similar result and suggested that the
gas was ejected from the galaxy before it had been sufficiently
processed by stars to increase the metal content, which could also be
expected for highly star-forming galaxies at $z\sim2$.

Given the gas we observe is $D=66$~kpc from the galaxy and has had to
travel through the CGM along the way, it is also likely that the
outflow has entrained more metal-poor material from accretion, past
outflows, and/or other CGM material and resulted in a lower
metallicity. Thus while the current metallicity might indicate the gas
is not completely new metal-rich material from the current star
formation activity, we can conclude that the gas had been in the
galaxy at some point in the past and was likely ejected via an
outflow. It is relevant to note that we assume a single low ionization
phase when inferring a metallicity, so we may be missing outflow mass
if metal-rich outflows are primarily hot, high ionization material
\citep[e.g.,][]{muzahid15}. The lower metallicity, low ionization gas
we observe would then be an indirect tracer of these outflows as
entrained material and provide more accurate information about the
kinematics and spatial distribution of the outflow than the
metallicity.

A troublesome aspect of inferring a total CGM metallicity is that
observations have not yet convincingly demonstrated that metallicity
is a good discriminator between accreting and outflowing gas for a
large sample of galaxies. The typical method for inferring a total
metallicity neglects the effects of dust depletion, as we have done
here. \citet{wendt20} found that minor axis sightlines have increased
amounts of dust, which would result in lower inferred metallicities if
dust is not accounted for, so our inferred value could be artificially
low. Additionally, the CGM is expected to be messy, with the various
baryon cycle processes mixing at all radii \citep[at least for
  $z<1$;][]{peroux16, pointon19iso, kacprzak19b}. In fact, simulations
often show that quasar sightlines pass through multiple structures
along the line-of-sight \citep{churchill15, kacprzak19a, peeples19},
making total CGM metallicities an insensitive probe of gas
flows. \citet{hafen19} suggested that the CGM becomes more well-mixed
towards lower redshift, further complicating the metallicities
measured observationally \citep[also see][]{hafen17}. For the
J1430$+$014 absorber studied here, we do see varying ionization
conditions (and likely metallicities) across the profile, particularly
when comparing {\SiII}, {\SiIII}, and {\SiIV} in the bluest
components. This further supports the discussion in the previous
paragraph that the observed gas may not be purely outflow material but
either represents an outflow that has swept up additional material on
its way through the CGM, or the nature of quasar sightlines means we
are probing a very large range in radius from the galaxy so that the
sightline passes through multiple separate structures with different
ionization conditions and metallicities. In contrast,
\citet{outflowsreview} suggest that large variations in metallicity
and ionization conditions between clouds across an absorption profile
might be a signature of outflows in addition to broad complex
kinematics spanning a few hundred {\kms}, kinematics that are not
consistent with rotation, $\alpha$-rich abundances, and dual-trough
absorption with weak absorption between. There are few studies thus
far with detailed metallicity and kinematic analyses in order to
investigate this further.

To improve the metallicity measurement and better determine the
structures along the line-of-sight, a multiphase and multicomponent
modeling approach would be best. Several works have done such a study
and found differing metallicities in each component and/or phase
\citep[e.g.,][]{tripp11, crighton15, crighton16, muzahid15, muzahid16,
  rosenwasser18, zahedy19LRG} and these insights could be used to
determine if outflows, inflows, tidal material, etc., are present
along the same sightline. In particular, a multiphase analysis would
provide insights into whether outflows are primarily traced by hot
high ionization gas with metallicities similar to the galaxy ISM
metallicity and whether the lower ions along the same sightline trace
cooler, more metal-poor gas entrained by the hot outflow, which is a
possible solution to the total metallicity problem suggested by
\citet{kacprzak19b} that might fit the scenario we observe here. The
difficulty with this method is that it requires partitioning each ion
between VP components and phases, and modeling a consistent set of VP
components across transitions with similar ionization phase. This is
particularly challenging when absorption is saturated and the
selection of lines is limited, i.e., the only {\HI} line observed is
saturated as in the system presented here. While much of our sample
has coverage of multiple {\HI} Lyman series lines and in some cases
the full series to properly do a multiphase and multicomponent
analysis, the J1430$+$014, $z=2.071$ absorber is not a good candidate
for this method. Without obtaining new spectra to better cover the
{\HI} Lyman series, we would need to make assumptions on how to
partition the saturated {\HI} between components and phases, which
would result in over-interpreting the absorption. Furthermore, the
vast majority of metallicities in the literature are total
metallicities. To study the evolution of the CGM from Cosmic Noon to
today, we have focused on total metallicities here.

Even though metallicity is a useful diagnostic for ruling out some
scenarios of gas flow origins, especially at higher redshift where CGM
mixing appears to be less efficient, a variety of properties such as
kinematics and galaxy morphologies must also be considered for a more
accurate picture, which we have done here. This approach has thus far
only been done a few times at Cosmic Noon.

\subsection{Other Gas Flow Interpretations}
\label{sec:othergasflows}

Other mechanisms and structures that could give rise to the observed
absorption include tidal streams \citep[e.g.,][]{yun94, chynoweth08,
  ggk1127, duc13, deblok18}, IGM accretion \citep[e.g.,][]{birnboim03,
  keres05}, an intragroup medium \citep[e.g.,][]{whiting06, ggk1127,
  bielby17, peroux17, pointon17, magiicat6, chen1127, hamanowicz20},
and intergalactic transfer \citep[e.g.,][]{angles17}.

\subsubsection{Tidal Stripping in a Major Merger}

While we have made the assumption throughout the paper that the galaxy
is a single, clumpy, edge-on disk, it has a non-zero probability of
being a major merger system in the late stages of merging (see
Section~\ref{sec:galmorph}). In this case, the absorbing gas observed
in the quasar spectrum could be tidal debris from the merging
process. Observations suggest that thin tidal streams extend out to
$\sim40$~kpc in the M81/M82 system \citep[e.g.,][]{yun94, deblok18},
$\sim80$~kpc in the Antennae galaxies \citep[NGC4038/9,
  e.g.,][]{schweizer78, hibbard01}, potentially beyond $\sim100$~kpc
in other systems/simulations \citep[see][for a review]{duc13}, and are
even visible out to at least $\sim25$~kpc in {\it HST} imaging and
MUSE observations of galaxies hosting CGM absorbers
\citep[Q1127$-$145, e.g.,][]{ggk1127, peroux17,
  chen1127}. \citet{hani18} studied the impact a simulated major
merger has on the CGM at $z<1$ by investigating the velocities, column
density profiles, covering fractions, and enrichment of the CGM
before, during, and after a merging event between two $\log
(M_{\ast}/M_{\odot})\sim10$ galaxies in the Illustris
simulations. They found that stellar tidal streams are not observed
beyond 50~kpc throughout the merging process, but that the velocity
field of the gas at large impact parameter supports an outflow
origin. The authors suggested that tidal stripping is not a primary
mechanism for distributing material into the CGM during a major
merger, at least beyond 50~kpc. This is further corroborated in the
FIRE simulations, where $z=2$ galaxies have gas flows that are
dominated by fresh accretion and wind recycling ($>70\%$), with a
minor contribution from mergers \citep{angles17}.

To better understand the contributions of tidal stripping and outflows
in a major merger, we estimated covering fractions of gas in the two
scenarios. In the first scenario, we assume that a major merger would
have tidal streams similar to those found in the Antennae galaxies,
with a radial extent of $\sim80$~kpc and a thickness of $\sim15$~kpc
\citep{schweizer78, hibbard01}. Assuming that there are two tidal
streams crossing a projected distance from a merging system of
$D\sim70$~kpc (the rough impact parameter of our absorber--galaxy pair
at $z=2.071$) and the streams themselves have a unity covering
fraction, we estimate a total tidal stream covering fraction at 70~kpc
of only $7\%$. This value could be larger if the tidal streams are
wider or cross the boundary multiple times (i.e., doubling the width
increases the covering fraction to $14\%$ and additionally having both
thicker streams cross the 70~kpc twice increases the value to $28\%$),
but could also be smaller if the streams are thinner, one or both of
the streams do not reach the impact parameter in question, and/or the
streams themselves have less than unity covering fraction. In the
second scenario, if we assume that bipolar outflowing gas has a unity
covering fraction within the outflow cones, then we find a total
outflow covering fraction at 70~kpc ranging between $16\%$ for a half
opening angle of $14^{\circ}$ and $83\%$ for a half opening angle of
$75^{\circ}$, where the half opening angles are the allowed bounds for
J1430$+$014, $z_{\rm gal}=2.0711$ from Table~\ref{tab:Mout}. While the
total outflow covering fraction could be smaller with a lower covering
fraction within the outflow itself and if some of the material does
not reach $D\sim70$~kpc, there is evidence that outflowing gas is not
well-collimated at $z\sim2$ \citep[e.g.,][]{steidel10, law12,
  nelson19}. This suggests that the total outflow covering fraction is
more likely to be on the higher side of the estimated range. Despite
the presence of both tidal streams and outflowing gas in the CGM
surrounding a major merger, the probability of intersecting outflowing
gas is much higher than the thin filamentary tidal material.

The simulations, the large impact parameter of the absorber--galaxy
pair studied here, the low covering fraction of tidal material, the
high covering fraction of outflowing material, and the similarity to
the outflows discovered in Makani \citep{rupkemakani} further suggest
that the absorption we measure in the CGM of the J1430$+$014, $z_{\rm
  gal}=2.0711$ galaxy is more likely outflowing material even if the
morphology is a major merger rather than an edge-on disk.

\subsubsection{IGM Accretion}

We rule out IGM accretion as the dominant source of the gas for
several different reasons. First, IGM accretion is expected to have a
low metallicity \citep{vandevoort+schaye12, hafen19} but the gas we
observe is relatively metal-enriched compared to the distribution of
$z=2-3$ LLSs. Second, accreting IGM filaments are expected to add
angular momentum to the host galaxy \citep{danovich12, danovich15},
which implies that the gas must align with the star forming disk and
be offset in velocity towards one side \citep{stewart11, stewart13,
  stewart17}. Recall that the gas studied here is located along the
minor axis of its host galaxy, where an accreting filament of gas is
unlikely to be directly perpendicular to the disk even at the somewhat
large impact parameter ($D=66$~kpc). Also, the gas spans both sides of
the systemic galaxy velocity, in contradiction to predictions and
major axis observations \citep{kacprzak11kin, ho17}. Third, accreting
filaments are expected to have small cross-sections, at least at low
redshift, and have yet to be directly observed. While it is possible
that accretion has larger cross-sections at Cosmic Noon when the
process is most active, the combination with the other two points
makes it unlikely that we have observed accreting material in the
first galaxy studied for our survey without specifically searching for
its signatures.

\subsubsection{Group Environments and Intergalactic Transfer}

It is possible that this absorber is located in a group environment
given the proximity of three additional galaxies in projection on the
sky in the {\it HST} image. However, we detect no obvious {\Lya}
emission, or any other lines consistent with a $z=2$ LBG, from these
three galaxies. The {\Lya} emission in the continuum galaxy spectrum
(bottom panels of Figure~\ref{fig:galspec}) appears to be
contamination from the host galaxy. This could be due to the seeing
(FWHM$\sim 1\arcsec$) and/or {\Lya} emission in the CGM of the
$z=2.0711$ host galaxy detailed here. At $z>3$ \citet{leclercq17}
studied the {\Lya} emission profiles around star-forming galaxies with
VLT/MUSE. The authors found that {\Lya} emission halos are ubiquitous
around star-forming galaxies, with $80\%$ of their sample of $\sim150$
galaxies having firmly measured {\Lya} emission extending out to scale
lengths of roughly $4-5$~kpc and even beyond 10~kpc \citep[also
  see][]{xue17}. The distance between our CGM host galaxy and the
non-related continuum galaxy with contaminating {\Lya} emission is
roughly 1\farcs8, corresponding to $D=15$~kpc, and is consistent with
the scale lengths measured.

At low redshift, group environments have a differential influence on
the various metal lines we observe in this system. Group environments
appear to enhance low ionization lines such as {\MgII} to a degree,
where the equivalent widths and radial extent are larger, and
kinematic spreads are comparable to outflows for small galaxy--galaxy
groups \citep[e.g.,][]{bordoloi11, magiicat6}. The strongest {\MgII}
absorbers may also be located in group environments \citep{nestor07,
  nestor11, gauthier13}. For more dense group environments and cluster
environments, a dearth of low equivalent width {\MgII} absorbers and a
smaller radial extent of the gas indicates that the cluster
environments are hot enough to ionize the gas to higher phases
\citep[e.g.,][]{lopez08, padilla09, andrews13}. The {\MgII} observed
here is fairly weak, but its kinematics still appear to be consistent
with outflows. In contrast, the intermediate and higher ions such as
{\CIV} and {\OVI} are more easily ionized to higher states in group
environments \citep[e.g.,][]{burchett16, burchett18, pointon17, ng19},
where the detection rate drops with an increasing density of the
environment. Given that the {\CIV} observed in this system is strong
and the {\MgII} is not enhanced, a group environment seems less
likely.

The CGM host galaxy is clearly a {\Lya} emitter, which is more likely
located in the distant outskirts of group environments (where group
interactions are minimal) or is isolated \citep{cooke13}. While this
may rule out nearby large galaxies, it does not rule out the
possibility that this gas is falling onto the host galaxy from a
satellite experiencing active star formation. Given the redshift and
the large KCWI spaxels ($\sim2-6$~kpc), we cannot detect satellite
galaxies surrounding the bright host. It is possible that satellites
may be detected by examining the {\Lya} emission halo in detail, which
may be clumpy and asymmetric when strong {\Lya}-emitting satellites
are present \citep{leclercq17}. Unfortunately our KCWI data are not of
sufficiently high spatial resolution to investigate this
further. Despite all this, it is less likely that this scenario
dominates the absorption signal due to the small cross-sections
expected \citep{gauthier10, martin12, tumlinson13}. The transfer of
material between galaxies in this scenario is expected to dominate
over other gas flows only at lower redshift, with only $<10\%$ of gas
flows at $z=2$ coming from intergalactic transfer in the FIRE
simulations \citep[][]{angles17}. The galaxies in these simulations
are instead dominated by wind recycling and fresh accretion at $z=2$,
which comprise $>70\%$ of gas flows.

\vspace{0.2in}

Under the assumption that the object we observe with {\Lya} is a
single, clumpy, edge-on galaxy on the SFR main sequence, the
scientific evidence best supports an interpretation that the observed
absorption is an outflow from an isolated galaxy at Cosmic Noon that
has likely swept up more metal-poor CGM material as it moves away from
the galaxy. With only a few $z=2-3$ absorber--galaxy pairs with galaxy
morphologies and CGM metallicities, it is difficult to make any
conclusions about the general properties of the CGM or outflows during
this epoch. However, we aim to build a sample of $\sim50$ {\MgII}
absorber--galaxy pairs with similar quality data and with coverage of
other ions for comparison to the low redshift, multiphase CGM.

\section{Summary and Conclusions}
\label{sec:summary}

We presented the first results from our CGM at Cosmic Noon with KCWI
program: minor axis gas that is likely outflowing and sweeping up
additional more metal-poor CGM material from a $z_{\rm gal}=2.0711$
star-forming galaxy in quasar field J143040$+$014939. The program was
designed by identifying quasar fields with {\MgII} absorbers at $1.9 <
z_{\rm abs} < 2.6$ in Keck/HIRES or VLT/UVES spectra and required that
each field have existing {\it HST} (ACS, WFPC2, and/or WFC3) or
Keck/NIRC2-LGSAO imaging for galaxy morphology modeling. We then
observed each field with KCWI to search for the host galaxies in
{\Lya} emission. From an analysis on a single absorber--galaxy pair in
the J1430$+$014 field, our results include the following:

\begin{enumerate}
  \item We identified the absorbing host galaxy with {\Lya} emission
    at $z_{\rm gal}=2.0711\pm0.0003$ in the KCWI datacube $D=66$~kpc
    away from the quasar sightline. The galaxy is isolated and
    star-forming with ${\rm SFR}_{\rm
      FUV}=37.8$~M$_{\odot}$~yr$^{-1}$.  It has a clumpy morphology,
    which is typical of $z=2-3$ star-forming galaxies, and we assume
    an edge-on ($i={85^{\circ}}_{-2}^{+5}$) inclination based on the
    object's small size, small clump separation, and main sequence
    SFR. A major merger morphology is also possible, but less likely
    given these characteristics. The quasar sightline probes the
    galaxy along the projected minor axis
    ($\Phi={89^{\circ}}_{-5}^{+1}$). The {\Lya} emission halo
    surrounding the galaxy appears to extend out to at least 1\farcs8,
    corresponding to $D\sim15$~kpc.

 \item The absorption system is roughly centered at the host galaxy
   systemic velocity, with an average offset of $\Delta v_{\rm
     gal-abs}=-27$~{\kms}. The full {\MgII} velocity spread of $\Delta
   v\sim210$~{\kms} means that the absorption spans both sides of the
   galaxy systemic velocity. The absorption is multiphase, with
   detected {\Lya}, {\MgII}, {\SiII}, {\SiIII}, {\SiIV}, {\CII}, and
   {\CIV}, where {\Lya} and the higher ions have larger velocity
   spreads. We constrain the {\HI} column density from only the
   saturated {\Lya} absorption line to the range $15.00\leq\log
   N_{\tiny \HI}\leq18.18$~{\cmsq}, indicating that this is a
   pLLS/LLS. The {\MgII} equivalent width,
   $W_r(2796)=0.24\pm0.01$~{\AA}, places this absorber--galaxy pair in
   line with the $W_r(2796)-D$ anti-correlation and within
   uncertainties of the log-linear fit to the anti-correlation from
   \citet{magiicat2, magiicat1} at $z<1$. For this single absorber,
   this indicates that there is no clear evolution of the CGM extent
   from low redshift.

 \item From photoionization modeling, the total CGM metallicity is
   inferred to be [Si/H]$=-1.5_{-0.3}^{+0.4}$, which is slightly more
   metal-rich than the peak in the distribution of pLLSs/LLSs
   metallicities at $z=2-3$, but is lower than the metallicities
   expected for purely outflowing gas at this epoch. Assuming the gas
   is an outflow, this lower metallicity is not unexpected since
   outflows likely entrain more metal-poor CGM gas from past outflows
   on their way out from the galaxy, especially to $D=66$~kpc. The
   ionization conditions, and therefore the metallicities, appear to
   vary across the absorption profile, which may be a signature of
   outflows, but the lack of additional {\HI} Lyman series lines for
   this system prevents further robust analyses to investigate
   this. Removing higher ionization ions such as {\CIV} and {\NV} from
   the photoionization modeling to account for possible multiphase
   structure does not significantly change the inferred value, but
   does suggest the true value could be 0.3~dex more metal-poor.

 \item The metal line absorption kinematics are comparable to those
   found in the Milky Way Fermi bubbles, where the bulk of the
   absorption is centered on $z_{\rm gal}$ with roughly symmetric
   higher velocity components \citep{fox15}. Under the assumption that
   the host galaxy has an edge-on inclination, modeling the {\MgII}
   kinematics as a biconical outflow suggests that the gas is a
   decelerating outflow, where we constrain the average radial outflow
   velocities over the range of $V_{\rm out}\sim109-588$~{\kms} for
   half opening angles of
   $\theta_0\sim14^{\circ}-75^{\circ}$. Assuming a constant outflow
   velocity, it would take on average $\sim111-597$~Myr for the gas to
   reach 66~kpc, corresponding to gas ejection redshifts between
   $z_{\rm eject}\sim2.15-2.55$. This is enough time for the galaxy to
   evolve from higher SFRs when the observed gas was ejected in an
   outflow to the SFR we measure at $D=66$~kpc and $z_{\rm
     gal}=2.0711$.

 \item Assuming the gas is outflowing, we measure mass outflow rates
   of $\dot{M}_{\rm out}\lesssim 52\pm37$~M$_{\odot}$~yr$^{-1}$. Given
   this range, the mass loading factors are $\eta \lesssim
   1.4\pm1.0$. These values are comparable to simulated galaxies of
   similar magnitude. They are also roughly an order of magnitude
   larger than those measured at $z\sim1$ with the MEGAFLOW survey
   \citep{schroetter16, schroetter19} but are comparable to those
   measured in an extreme metal-rich outflow at $z=0.4$
   \citep{muzahid15}. The outflow from this galaxy at Cosmic Noon may
   be more energetic and more efficient than those at lower redshifts
   but this is not yet well-constrained.

 \item Other gas flow origins such as tidal streams, IGM accretion,
   and an intragroup medium with intergalactic transfer do not appear
   to explain the observed properties of the absorption and its
   relation to the host galaxy as well as an outflow origin, nor are
   they as probable. While an outflow best fits the gas kinematics and
   quasar--galaxy orientation, the outflow has likely swept up
   additional more metal-poor material from accretion and/or past
   outflows on its way through the CGM.

 \item We caution against making strong interpretations as to the
   origin of CGM material based solely on total metallicity with the
   data and analyses available since lower redshift work has not yet
   convincingly demonstrated that total metallicities separate out
   accretion and outflows. Our outflow interpretation here was
   determined based on evidence that the galaxy is on the star
   formation rate main sequence, has an edge-on inclination, the
   quasar probes the galaxy projected minor axis, the broad absorber
   kinematics that are roughly symmetrically centered on $z_{\rm
     gal}$, and more cautiously, a total metallicity in the upper half
   of the pLLS/LLS distribution. A multiphase and multicomponent
   metallicity analysis may be more robust for determining whether the
   ionization structure across the absorption profile is indicative of
   outflows or multiple structures along the quasar sightline, but
   this is not possible here due to the lack of additional
   non-saturated {\HI} transitions.

\end{enumerate}

These first results show how powerful the combination of
high-resolution quasar spectra, {\it HST} images, and KCWI integral
field spectroscopy is for studying the CGM at Cosmic Noon. We have
examined the gas flow properties in as much detail as has been done at
$z<1$ for a single absorber--galaxy pair and this is one of only a few
systems at this epoch with this wealth of data, specifically including
galaxy morphologies and CGM metallicities. In the future, we plan to
compile a sample of roughly 50 $z=2-3$ absorber--galaxy pairs for not
only detailed single system studies, but also ensemble studies, which
can then be compared to those at low redshift to study how gas flows
evolve over ten billion years.

\acknowledgments

We thank the anonymous referees for comments that improved this
manuscript. N.M.N.~thanks L.~Rizzi, G.~Doppmann, J.~O'Meara,
A.~Ferr\'{e}-Mateu, S.~Cantalupo, and D.~Rupke for useful KCWI
discussions, and J.~Cooke, T.~Yuan, and D.~B.~Fisher for high redshift
galaxy discussions.

N.M.N., G.G.K., and M.T.M.~acknowledge the support of the Australian
Research Council through {\it Discovery Project} grant
DP170103470. Parts of this research were supported by the Australian
Research Council Centre of Excellence for All Sky Astrophysics in 3
Dimensions (ASTRO 3D), through project number CE170100013. C.W.C. was
supported by the National Science Foundation through grant NSF
AST-1517816 and by NASA through HST grant GO-13398 from the Space
Telescope Science Institute, which is operated by the Association of
Universities for Research in Astronomy, Inc., under NASA contract
NAS5-26555.

Some of the data presented herein were obtained at the W. M. Keck
Observatory, which is operated as a scientific partnership among the
California Institute of Technology, the University of California and
the National Aeronautics and Space Administration. The Observatory was
made possible by the generous financial support of the W. M. Keck
Foundation. Observations were supported by Swinburne Keck programs
with KCWI: 2017B\_W270, 2018A\_W185, 2018B\_W232; with NIRC2-LGSAO:
2019B\_W237; and NASA Keck program with KCWI: 2020B\_N021. The authors
wish to recognize and acknowledge the very significant cultural role
and reverence that the summit of Maunakea has always had within the
indigenous Hawaiian community. We are most fortunate to have the
opportunity to conduct observations from this mountain.

\facilities{Keck:II (KCWI), VLT:Kueyen (UVES), {\it HST} (ACS)}

\software{Astropy \citep{astropy}, Cloudy \citep{cloudy}, Matplotlib
  \citep{matplotlib}, numpy \citep{numpy20}, PyAstronomy, scipy
  \citep{scipy}, qfitsview, VPFIT \citep{vpfit}}

\bibliography{refs}
\bibliographystyle{aasjournal-hyperref}{}

\end{document}